\documentclass[fleqn,a4paper,12pt]{article}
\usepackage{amssymb}

\usepackage{amsmath}
\usepackage[dvips]{graphicx}
\usepackage{epsfig}

\newcommand{\bab}{\end{gather}}
\newcommand{\ri}{{\mathrm i}}
\newcommand{\p}{\partial}
\newcommand{\bea}{\begin{array}}
\newcommand{\eea}{\end{array}}
\newcommand{\beg}{\begin{gather}}

\catcode`@=11 \long
\def\@caption#1[#2]#3{\par\addcontentsline{\csname
ext@#1\endcsname}{#1} {\protect\numberline{\csname
the#1\endcsname}{\ignorespaces #2}} \begingroup \small
\@parboxrestore \@makecaption{\csname fnum@#1\endcsname}
{\ignorespaces #3}\par \endgroup} \catcode`@=12

\newcommand{\la}{\label}

\catcode`@=11 \long
\def\@caption#1[#2]#3{\par\addcontentsline{\csname
ext@#1\endcsname}{#1} {\protect\numberline{\csname
the#1\endcsname}{\ignorespaces #2}} \begingroup \small
\@parboxrestore \@makecaption{\csname fnum@#1\endcsname}
{\ignorespaces #3}\par \endgroup} \catcode`@=12

\begin{document}

\allowdisplaybreaks
 \begin{titlepage} \vskip 2cm

\begin{center} {\Large\bf Integrable and superintegrable quantum mechanical systems with position dependent
masses invariant with respect to one parametric Lie groups. 2. Systems with dilatation and shift symmetries.}

 \vskip 3cm {\bf {A. G. Nikitin }\footnote{E-mail:
{\tt nikitin@imath.kiev.ua} }
\vskip 5pt {\sl Institute of Mathematics, National Academy of
Sciences of Ukraine,\\ 3 Tereshchenkivs'ka Street, Kyiv-4, Ukraine,
01024}}, and \\
Universit\'a del Piemonte Orientale,\\
Dipartimento di Scienze e Innovazione Tecnologica,\\
viale T. Michel 11, 15121 Alessandria, Italy\end{center}
\vskip .5cm \rm
\begin{abstract} { 3d  quantum mechanical
systems with position dependent masses (PDM) admitting at least one second order integral of motion and
symmetries with respect to dilatation or shift transformations are classified. Twenty seven such systems are
specified and the completeness of the classification results  is proved. In this way the next step to the
complete classification of integrable PDM system is realized. }
\end{abstract}
\end{titlepage}
\section{Introduction\label{int}}

 Symmetry is one of the most fundamental concept of  quantum mechanics, and it is natural that just the symmetry lies in the focus of this research field.
A systematic search for Lie symmetries of Schr\"odinger equation started  in papers \cite{Hag, Nied,
And} and \cite{Boy} where the maximal invariance groups of this equation with arbitrary scalar
potential have been discovered. For symmetries of this equation with scalar and vector potentials and
corrected results of classical paper \cite{Boy}  see  papers \cite{N22, N32}. Lie symmetries of
Schr\"odinger equation with matrix potentials and of quasirelativistic Schr\"odinger equation are classified in \cite{N42, N52}.

More general  symmetries, namely, the second order symmetry operators for 2$d$ and 3$d$
Schr\"odinger equation have been classified in \cite{wint1}, \cite{BM} and \cite{ev1}, \cite{ev2}.
 Just such symmetries characterize integrable and superintegrable systems
\cite{wint01}.

An interesting research field is formed by
superintegrable systems with spin.   The first example of such systems  presented  in
paper \cite{Pron1}, the  systematic study of them
 started with papers \cite{w6,  w8} where the systems with spin-orbit interaction were classified.
Superintegrable  systems with Pauli type interactions were studied
in \cite{N5, N1} and  \cite{N2}, the examples of relativistic systems can be found
 in \cite{N2} and \cite{N77}.

One more research field closely related to the discussed above are symmetries  of
Schr\"odinger equations with position dependent mass. This equation  is requested in many branches of modern theoretical physics, whose list can be found,
e.g., in   \cite{Rosas, NZ}.  These  symmetries are studied much less. Nevertheless  symmetries
with respect to the continuous groups have been
classified in papers \cite{NZ} -\cite{AGN}.

The situation with the higher symmetries of the PDM  systems is  more
complicated.
The 2d classical systems with position dependent mass which admit second order
integrals of motion  are perfectly classified  \cite{Koen, Kal1, Mi, Kal4}. In particular, the two dimensional
second-order (maximally) superintegrable systems for Euclidean 2-space had been classified also algebraic geometrically \cite{Kress}, see also \cite{Va} for the modern proof of the fundamental Koenig theorem \cite{Koen}.  On the other hand the more interesting for physics 3d systems and their higher symmetries are not completely classified jet. More exactly, the maximally superintegrable systems  (i.e., admitting the maximal possible number of integrals of motion) and (or) for the system whose integrals of motion are supposed to satisfy some special conditions like the functionally linearly dependence \cite{bern} are well studied  \cite{ Kal31}. The same is true for the so called nondegenerate
systems \cite{Cap, Vol}. A certain progress has been achieved  in the
classification of the  semidegenerate systems  \cite{Mil}.

Surely, the maximally superintegrable  and nondegenerate systems are both important and interesting. In particular, they admit solutions in multi separated coordinates \cite{Ra, Bala2, Rag1}. On the other hand, there are no reasons to ignore the PDM systems which admit second order integrals of motion but are not necessary either superintegrable or nondegenerate. And just such systems are the subject of our study.

The main stream  in studying of superintegrable systems with PDM is the investigation of classical
 Hamiltonian systems. Surely there exist the analogous  quantum mechanical systems which in
 principle can be obtained starting with the classical ones by the second quantization
 procedure. However, the mentioned procedure is not unique, and in general    it is possible to
 generate few inequivalent quantum systems which have the same classical limit. In addition, a part of
 symmetries and integrals  of motion can disappear in the classical limit $h\to 0$ \cite{Hit}. And it is the reason why we classify
 superintegrable quantum systems directly.

 In view of the complexity of the total classification problem of  integrals of motion for 3d quantum PDM systems   it is reasonable to separate it to well defined subproblems which can have their own  values. The set of such  subproblems can be treated as optimal one if solving them step by step we can obtain the complete classification.

We choose the optimal set of subproblems  in the following way.

It was shown in \cite{NZ} the PDM Schr\"odinger equation can admit one, two, three, four, or six
parametric Lie symmetry groups. In addition, there are also such equations which have no Lie
symmetry. In other words, there are six well defined classes of such equations which admit n-parametric Lie groups with $n=1, 2, 3, 4, 6$ or do not have any Lie symmetry. And it is a natural idea to search for second order integrals of
motion consequently for all these classes.

The systems admitting six- or four-parametric Lie groups are not too interesting since the related Hamiltonians cannot include non-trivial potentials. That is why we started our research with the case of three-parametric groups. The classification of the corresponding PDM systems admitting second order integrals of motion was obtained in  \cite{AG}. There were specified 38 inequivalent PDM systems together with their integrals of motion. The majority of them are new systems which are not not maximally superintagrable.

Notice that the superintegrable 3d  PDM systems invariant with respect to the 3d rotations
have been classified a bit earlier in paper \cite{154} where their supersymmetric aspects were discussed also. For relativistic aspects of superintegrability see \cite{N77} , \cite{AG3}.

The systems admitting two-parametric Lie groups and second order integrals of motion had been classified in \cite{AG1}. We again find a number of new systems in addition to the known maximally superintegrable ones.

The natural next step is to classify the systems which admit one parametric Lie groups and second order integrals of motion.
As it was shown in \cite{NZ}, up to equivalence there are six one parametric Lie groups which can be accepted by equation (\ref{se}). These groups are:
\begin{itemize}
\item{Rotations around the fixed axis};\item{Dilatations;}\item {Shifts along the fixed axis;}\item{Superposition of the rotations and shifts;}\item{superposition of the rotations and dilatation;}\item{Supeposition of the rotations and the special conformal transformations.}
\end{itemize}

The integrable and superintegrable PDM admitting the symmetry w.r.t. the mentioned rotations, i.e., possessing the cylindric symmetry had been classified in paper \cite{AG5}. The number of such systems appears to be rather extended. Namely, the 68 inequivalent systems admitting second order integrals of motion have been discovered in \cite{AG5}. The majority of these systems are neither maximally integrable nor degenerate and so are new and cannot be discovered using the approaches exploited in \cite{Kal1, Mi, Kal4}, \cite{Cap, Vol} and \cite{Mil}.

In the present paper  we make the next step to the complete classification of the integrable and
superintegrable 3d PDM systems admitting one parametric Lie symmetry groups. Namely, we
present  the complete classification of two subclasses of such systems. The first class includes the systems
possessing the dilatation symmetry while the system belonging to the second class are invariant w.r.t. the
shifts along the third coordinate axis. Just these symmetries are presented in Items 3 and 4. Formally speaking  the results of paper \cite{AG5} and of the present paper cover the half part of the possible one parametric symmetries enumerated in the above presented items. But in fact these results include the main part of  the systems admitting these symmetries. The number of the  systems presented in any of these  papers is more extended that the total number of the systems admitting the symmetries enumerated in the three last items.

Notice that    the second order integrals of motion which can be admitted by the PDM systems invariant with respect to  the two parametric Lie group including both the mentioned shifts and dilatation transformations have been classified in \cite{AG1} and it is interesting to extend this result to the case of the systems invariant w.r.t. its one dimensional subgroups. In addition,
the second order integrals of motion of such systems have the following unique property: they are generated by the specific   conformal Killing tensors whose traceless parts are polynomials of  order  less than three, and this one more reason   to present both the mention subclasses in one  paper.

To classify the mentioned systems we use the optimised  algorithm of solution of the related
 determining equations proposed in \cite{AG5}, see Section 2 in the following text.

The number of the inequivalent systems with the dilatation symmetry appears to be equal to ten, including five integrable, three superintegrable and two maximally superintegrable ones. Notice that a special  restricted class of such systems was studied earlier in papers \cite{AG} and \cite{AGG}.

We present also eighteen inequivalent systems which admit second order integrals of motion and are invariant with respect to the shift transformations. Among them there are seven integrable, seven superintegrable and four maximally superintegrable ones. This means that the majority of the found systems is new.

The paper is organized in the following manner. In addition to Introduction and Discussion sections
it includes three large parts  named as Sections 2, 3 and 4.

 In Sections 2 we present the general discussion which includes the basic definitions (Section 2.1),
  the deduction of the determining equations (Section 2.2)
 and their   optimization (Section 2.3). In Section 2.4 we formulate  the  algorithm for the construction
 of general solutions of the determining equations which is used in the present and our previous papers.

Rather extended Sections 3 and 4 include the solutions of the classification problems formulated in
the above.

In Section 3 we classify the PDM systems which are invariant with respect to the dilatation
transformations and admit second order integrals of motion. In subsection 3.1 we show that the
related determining equations can be essentially simplified and decoupled to the subsystems which
can be exactly solved. In subsections 3.2-3.4 we find and present the solutions of these subsystems
together with the corresponding integrals of motion.

In Section 4 the shift invariant PDM systems are classified. In the beginning of this section we
discuss the specific properties of the related integrals of motion and the ways to use these properties to
optimise the solution procedure. The simplified determining equations are solved in subsection  4.1
 which makes it possible to classify the systems admitting at least one second order integral of
motion. The cases when the related systems are superintegrable and maximally superintegrable are
specified in Subsection 4.2.

Both Sections 3 and 4 are ended by the classification tables when the obtained results are summarized.
The last Section 5 includes the discussion of the obtained results.
\section{Formulation of the problem and algorithm for its solution}
\subsection{Basic definitions}
We will  search for superintegrable  stationary  Schr\"odinger equations with position dependent mass of
the following generic form:
\begin{gather}\la{se}
   \hat H \psi=E \psi,
\end{gather}
where
\begin{gather}\la{A1}  H=\frac12(M^\alpha p_a M^{\beta}p_aM^\gamma+
M^\gamma p_a M^{\beta}p_aM^\alpha)+ \hat V(x)\end{gather}
where $p_a=-i\p_a$, $M=M({\bf x})$ is a function of spatial variables ${\bf x}=(x_1,x_2,x_3)$, associated with
the position dependent mass, $\alpha, \beta$  and $\gamma$ are the so called
ambiguity parameters satisfying the condition $\alpha+\beta+\gamma=-1$, and and summation from 1 to 3 is
imposed over the repeating index $a$ (i.e.,  Einstein’s summation convention is used).
We will use the summation convention for all repeating indices not necessary for one up and second down.

There are various physical speculations how to fix the ambiguity parameters in the particular models based on
equation (\ref{A1}) . However, the systems with different values of these parameters are mathematically
equivalent up to redefinition of potentials $V({\bf x})$. To obtain the most simple and compact form of the
related potentials
 we will fix these parameters in
the following manner: $\beta=0, \gamma=\alpha=-\frac12$ and denote $f=f({\bf  x})=\frac1{M({\bf x})}$.
As a result hamiltonian (\ref{A1}) is reduced to the following form
 \begin{gather}\la{H} H=f^\frac12p_ap_af^\frac12+ V({\bf x})\end{gather}
where $p_ap_a=p_1^2+p_2^2+p_3^2$.
 In paper \cite{AG5} we find all inequivalent PDM systems invariant with respect to the mentioned rotations, which admit second order integrals of motion, i.e.,  differential operators of the following generic form
 \begin{equation}\label{Q}
    Q=\p_a\mu^{ab}\p_b+\eta
\end{equation}
commuting with Hamiltonian (\ref{H}). The number of such systems appears to be rather extended. Namely, we specified  68 versions of them defined up to arbitrary functions and arbitrary parameters.

In the present paper we search for the integrable PDM systems which are invariant with respect to the dilatation and shift transformations. The generic form of the related
 inverse masses $f$ and potentials $V$ can be represented by  the following formulae \cite{NZ}:
\begin{gather}\la{f_V1}f=\frac{r^2}{F(\varphi,\theta)}, \quad V=V(\varphi,\theta)\end{gather}
for the case of dilatation symmetry and
\begin{gather}\la{f_V2}f=F(x_1,x_2), \quad V=V(x_1,x_2)\end{gather}
for symmetries with respect to the shifts along the third coordinate axis,
where $F(.)$ and $V(.)$ are arbitrary functions whose arguments are fixed in the brackets,
\begin{gather*}   r^2=x^2_1+x_2^2+x_3^2,\quad \varphi=\arctan\left(\frac{x_2}{x_1}\right), \ \ \theta=\arctan\left(\frac{\tilde r}{x_3}\right).\end{gather*}

By definition, operators  $Q$ should commute with $H$:
\begin{equation}\label{HQ}[ H,Q]\equiv  H Q-Q H=0.\end{equation}

Our task is to find all inequivalent PDM Hamiltonians  with specific arbitrary elements $f$ and $V$ whose generic form is fixed in (\ref{f_V1}) and (\ref{f_V2}) which admit at least one integral of motion (\ref{Q}) satisfying (\ref{HQ}). As it was noted in \cite{154} it is reasonable to use  representation (\ref{A1}) for the hamiltonian since just this representation leads to the most simple form of the determining equations for symmetries. And this is why we will use it in the following calculations.
\subsection{Determining equations}

Evaluating the commutator in (\ref{HQ})
and equating to zero the coefficients for the linearly independent differential operators $\p_a\p_b\p_c$ and $\p_a$ we come to the following determining equations for arbitrary elements $f, V$ of the Hamiltonian and functions $\mu^{ab}$ , $\eta$ defining integrals of motion (\ref{Q}) \cite{AG1}:
\begin{gather}\la{m0}5\left(\mu^{ab}_c+\mu^{ac}_b+ \mu^{bc}_a\right)=
\delta^{ab}\left(\mu^{nn}_c+2\mu^{cn}_n\right)+
\delta^{bc}\left(\mu^{nn}_a+2\mu^{an}_n\right)+\delta^{ac}
\left(\mu^{nn}_b+2\mu^{bn}_n\right),\\
\la{m1}
 \left(\mu^{nn}_a+2\mu^{na}_n\right)f-
5\mu^{an}f_n=0,\\\la{m2}\mu^{an}V_n-f\eta_a=0\end{gather}
  where $\delta^{bc}$ is the Kronecker delta, $f_n=\frac{\p f}{\p x_n}, \
  \mu^{an}_n=\frac{\p \mu^{an}}{\p x_n }$, etc., and summation is imposed over the repeating indices
  $n$ over the values $n=1,2,3$, i.e., $\mu^{an}f_n=\mu^{a1}f_1+\mu^{a2}f_2+\mu^{a3}f_3$ and
  $\mu^{an}V_n=\mu^{a1}V_1+\mu^{a2}V_2+\mu^{a3}V_3$.

\subsection{Linearization of the determining equations}

Equation (\ref{m0}) defines the conformal Killing tensor. Its particular
solution  is $\mu^{ab}=\mu_0^{ab}$ where
\begin{gather}\la{K0}\mu_0^{ab}=\delta^{ab}g({\bf x})\end{gather}
 with arbitrary function $g({\bf x})$. In addition the conformal Killing tensors  include fourth order polynomials in $\bf x$ which are given by the following formulae:
 of  the following form (see, e.g., \cite{Kil}) :
\begin{gather}\la{K1}
\mu^{ab}_1=\lambda_1^{ab},\\
\la{K2}
\mu^{ab}_2=\lambda_2^a x_b+\lambda_2^b x_a,
\\\la{K3}
\mu^{ab}_3=(\varepsilon^{acd}\lambda_3^{cb}+ \varepsilon^{bcd}
\lambda_3^{ca})x_d,\\
\la{K4}
\mu^{ab}_4=(x_a\varepsilon^{bcd}+x_b\varepsilon^{acd}) x_c\lambda^d_4,
\\\la{K5}\mu^{ab}_5=
k x_ax_b,\\\la{K6}
\mu^{ab}_6=\lambda_6^{ab}r^2-(x_a\lambda_6^{bc}+x_b\lambda_6^{ac})x_c
,\\
\la{K7}\mu^{ab}_7=(x_a\lambda_7^b+x_b\lambda_7^a)r^2-4x_ax_b\lambda_7^c x_c,
\\\la{K8}\mu^{ab}_8= 2(x_a\varepsilon^{bcd} +x_b\varepsilon^{acd})
\lambda_8^{dn}x_cx_n- (\varepsilon^{ack}\lambda_8^{bk}+
\varepsilon^{bck}\lambda_8^{ak})x_cr^2,\\
\mu^{ab}_9=\lambda_9^{ab}r^4-2(x_a\lambda_9^{bc}+x_b\lambda_9^{ac})x_cr^2+
(4x_ax_b+\delta^{ab}r^2)\lambda_{9}^{cd}x_cx_d\la{K9}
\end{gather}
where $r=\sqrt{x_1^2+x_2^2+x_3^2}$, $\lambda_n^a $ and $\lambda_n^{ab}$  are
arbitrary parameters, satisfying the conditions $\lambda_n^{ab}=\lambda_n^{ba}, \lambda_n^{bb}=0$.

 Thus we can represent the generic integral of motion (\ref{Q})
 in the following form:
 \begin{gather}\la{Q0} Q=Q_{(0)}+\sum Q_{(n)}+\eta\end{gather}
 where
 \begin{gather}\la{Q1}Q_{(0)} =P_a g({\bf x}) P_a, \ Q_{(n)}=P_a\mu_m^{ab}P_b, \ m=1, 2, .. 9.\end{gather}

We have in hands the generic solution of the first subsystem of the determining equations presented in (\ref{m0}). The remaining determining equations, i.e.,  (\ref{m1}) and (\ref{m2})
represent the coupled system of three {\it nonlinear} partial differential equation equations for
two unknowns $g({\bf x})$ and $f({\bf x})$. However this systems can be linearized by introducing   the
new dependent variables $M$ and $N$ connected with $f$ and $g$ in the following manner:
\begin{gather}\la{MN}f=\frac1M, \ g=\frac{ V}M.\end{gather}
As a result we transform subsystem (\ref{m1}) to the following form:
\begin{gather}\la{mek}
 \left(\mu^{nn}_a+2\mu^{na}_n\right)M+
5(\mu^{an}M_n+N_a)=0.\end{gather}

 Equation (\ref{m2}) in its turn
 can be effectively simplified by introducing the  new dependent variables
 $\tilde M$ and $\tilde N$ in the following way:
 \begin{gather}\la{MN!}V=\frac{\tilde M}{M},\ \eta=\frac{N\tilde M}M-\tilde N\end{gather}
 which reduce (\ref{m2}) to the following form:
 \begin{gather}\la{me2}
 \left(\mu^{nn}_a+2\mu^{na}_n\right)\tilde M+
5(\mu^{an}\tilde M_n+\tilde N_a)=0\end{gather}
which is analogous to (\ref{mek}).

Just the linearised determining equation (\ref{mek}) will be used in the
following to solve our classification problem. Solving this equation we will find  the mass functions of the PDM
systems admitting second order integrals of motion of generic form  (\ref{Q0}). In fact in this way we will
immediately
find
the corresponding potential also since  equation (\ref{me2}) coincides with  (\ref{mek}).
\subsection{The algorithm for searching of general solution of the determining equations}
Equation (\ref{mek}) looks rather gentle but in fact we are supposed to deal with a very
complicated system of the coupled partial differential equations, including three dependent and three
independent variables. This complexity is caused by the fact  that the conformal Killing tensor $\mu^{ab}$
present in
 (\ref{mek}) includes as many as 35 arbitrary parameters, refer to (\ref{K1})-(\ref{K9}).
 However, for the case of the PDM systems admitting at least one parametric Lie symmetry groups this system can be effectively simplified and decoupled to the subsystems with the essentially reduced number of these parameters.

The algorithm we use to classify the PDM systems admitting second order integrals of motion and one parametric Lie groups includes the following steps.

1. To deduce the determining equations for the variable coefficients of the integrals of motion.
This step was represented in the beginning of  Section 2.

2. To linearize the determining equations. It has been done in the previous section where the linear versions of the determining equations were obtained, refer to equations (\ref{mek}) and (\ref{me2}).

3. To solve step by step equations (\ref{mek})  for all the cases when the PDM system
admits one of the symmetry groups enumerated in the  items present in the Introduction.
Notice that  in any of the mentioned cases the determining equations are much more simple than in the
generic case since the arbitrary elements $f$ and $V$ are essentially restricted by the symmetry condition.
For example, in the case of the complete rotation invariance of the considered PDM system these arbitrary
elements  should be functions of the only (radial) variable, and the determining equations are reduced to
the ordinary differential ones \cite{AG}.

4. In addition to the previous note, for the cases when a Lie symmetry is present  the systems of the determining
equations can be effectively decoupled to the relatively simple subsystems which can be  solved exactly.
Just such decoupling is the next step of our algorithm.

The ways to the mentioned decoupling are strongly dependent on the postulated   Lie symmetries.
In the case of symmetry with respect to rotations
 the decoupled  subsystems
describe separately the scalar, vector and tensor integrals of motion with different fixed parities \cite{AG5}. There
are also specific ways to the decoupling for the cases of dilatation and shifts invariance which are presented in the
following text, see Section 3.2 and the beginning of Section 4

5. The effective tools for the additional simplification of the determining equations are presented by the
continuous and discrete equivalence transformation. For the systems analysed in the present paper
such transformations are discussed in Section 2.3.

6. The last step of the algorithm is to find the generic solutions of the decoupled and simplified
subsystems taking the care on the possible presence of special solutions. It is important to stress that this
step is by no means trivial or easy since the simplified subsystems are still the systems of coupled partial
 differential equations with variable coefficients. However, using special and some times rather sophisticated
 approaches we were able to find the related exact general solutions in explicit form.

In the following sections we apply steps 4-6 of the presented algorithm for the classification of PDM systems
admitting second order integrals of motion and one parametric Lie groups specified in Items 2 and 3 presented
in Introduction, i.e, groups of dilatation and shift transformations. The cases of the remaining groups specified
in Items 4-6 will be considered in the following (and final) paper of this series.

\section{PDM systems invariant with respect to dilatation transformations}
In this section we solve the first part of our classification problem and consider the systems which are invariant with respect to the one parametric Lie group of dilatation transformations. The corresponding determining equations  can be effectively decoupled and solved.
\subsection{Equivalence relations}

Any classification problem is solved up to equivalence relations, and we need a clear definition of them in our particular case.

    Non degenerated changes of dependent and independent variables of  equations  (\ref{se}), (\ref{A1}) are equivalence transformations provided they keep their  generic form up to the explicit expressions  of  arbitrary elements $f$  and $V$.  They have a structure of a continuous group which is extended by  some discrete elements.

In accordance with \cite{NZ} the maximal continuous equivalence group of equation  (\ref{se}) is  the group of conformal transformations of the 3d Euclidean space which we denote as C(3). The corresponding Lie algebra c(3) is a linear span of the following differential operators:
\begin{gather}\label{QQ}\begin{split}&
 P^{a}=p^{a}=-i\frac{\partial}{\partial x_{a}},\quad L^{a}=\varepsilon^{abc}x^bp^c, \\&
D=x_n p^n-\frac{3\ri}2,\quad K^{a}=r^2 p^a -2x^aD\end{split}
\end{gather}
where $r^2=x_1^2+x_2^2+x_3^2$. Operators $ P^{a},$ $ L^{a},$ $D$ and $ K^{a}$ generate shifts, rotations, dilatations and pure conformal transformations respectively.

In addition, equation (\ref{se}) is form invariant with respect to the reflections of spatial variables and exactly invariant w.r.t. the  following discrete transformations:
 \begin{gather}\la{IT} x_a\to
\tilde x_a=\frac{x_a}{r^2},\quad \psi({\bf x})\to \tilde x^3\psi(\tilde{\bf x})\end{gather}
where $\tilde x=\sqrt{\tilde x_1^2+\tilde x_2^2+\tilde x_3^2}.$

The presented speculations are valid for an abstract system  (\ref{se}). However in our case the system should be invariant w.r.t. the conformal transformations, and to keep this property we have to reduce the equivalence group to its subgroup which does not affect the generator of these transformation, i.e., $D$. In other words, the equivalence algebra whose base is presented in (\ref{QQ}) is reduced to its subalgebra whose elements commute with $D$. And this subalgebra is nothing but so(3) $\oplus e_1$ spanned on $L_1, L_2, L_3$ and $D$. In addition, the mentioned discrete transformations are valid too.

\subsection{Decoupling of the determining equations}

Formulae (\ref{K0}) and (\ref{K1})-(\ref{K9}) include an arbitrary function and 35 arbitrary parameters, so
the system of determining equations (\ref{mek}) is rather  complicated. Happily   the considered  PDM systems by definition should be invariant w.r.t. the scaling transformations. It means that the related equations (\ref{mek}) cannot include
linear combinations of all polynomials listed in  (\ref{K1})-(\ref{K9})  but only homogeneous
ones.  Indeed, integral of motion (\ref{Q}) including inhomogeneous terms by construction is not invariant w.r.t. the mentioned transformations and so can be transformed to linear combination of integrals whose coefficients are homogeneous polynomials with arbitrary multipliers. In other words, the determining equations     (\ref{mek})
 are reduced to the five  decoupled subsystems corresponding to the Killing  tensors
which are $n$-order homogeneous polynomials with $n=0, 1, 2, 3, 4$. Moreover, since our Hamiltonians
(\ref{H}) with arbitrary elements
(\ref{f_V1}) are invariant with respect to the inversion transformation (\ref{IT})
 we can restrict ourselves to the polynomials of order $n < 3$, since symmetries with $n$=3 and
 $n$=4 appears to be equivalent to ones with $n=1$ and $n=0$ correspondingly. The corresponding
 Killing tensors  are $\mu_0^{ab}+\mu_1^{ab}, \mu_0^{ab}+\mu_2^{ab}+\mu_3^{ab}$ and
 $ \mu_0^{ab}+ \mu_5^{ab}+\mu_6^{ab}$ which we represent in (\ref{K0}) and the following formulae:
\begin{gather}\la{mm0}\mu^{ab}=\lambda^{ab}+\delta^{ab}g({\bf r}),\\\la{mm1}
\mu^{ab}=\lambda^a x_b+\lambda^b x_a-(2-\kappa)\delta^{ab}\lambda^cx_c+(\varepsilon^{acd}\lambda^{cb}+ \varepsilon^{bcd}
\lambda^{ca})x_d+\delta^{ab}g({\bf r}),\\\la{mm2}
\mu^{ab}=\mu_{(1)}^{ab}+\mu_{(2)}^{ab}+\delta^{ab}(g({\bf r})\end{gather}
where
\begin{gather}\la{mm3}\begin{split}&\mu_{(1)}^{ab}=(x_a\varepsilon^{bcd}+
x_b\varepsilon^{acd})\lambda^dx_c,\\&\mu_{(2)}^{ab}=
\lambda^{ab}x^2-(x_a\lambda^{bc}+x_b\lambda^{ac})x_c+\delta^{ab}(g({\bf r})-2\lambda^{cd}x_cx_d).\end{split}\end{gather}

We omit the subindices present in the related formulae (\ref{K1})-(\ref{K7}) and  ignore the tensor $\mu_4^{ab}$ which corresponds to the squared dilatation operator being the transparent symmetry.

Notice that $g({\bf r}) $  should be homogeneous functions of independent variables. Moreover their homogeneity degree should be zero in  case (\ref{mm0}), one in case (\ref{mm1}) and two in case (\ref{mm2}).

To classify the scale invariant PDM systems admitting second order integrals of motion
it is sufficient to solve the determining equations   (\ref{MN}) where $\mu^{ab}$ are special Killing
tensors fixed in (\ref{mm1}),   (\ref{mm2}). In other words, our classification problem is decoupled to
three subproblems corresponding to symmetries whose differential terms are independent on $x_a$, are
linear in $x_a$ or are quadratic in these variables.

\subsection{Symmetries with differential terms $Q_{(n)}$ (\ref{Q1}) independent on spatial variables}
Consider step by step  all equations (\ref{MN}) including Killing tensors (\ref{mm0}),  (\ref{mm1}), and  (\ref{mm2}).

In the case presented in (\ref{mm0}) operator (\ref{Q}) and equations (\ref{MN})  take the following form
\begin{gather}\la{Q11}Q=P_a(\lambda^{ab}+\delta^{ab}g({\bf x}))P_b+\eta\end{gather} and
\begin{gather}\la{e1}\lambda^{ab}M_b+N_a=0.\end{gather}

Let us note that in fact we have two more equations additional to system (\ref{e1}). By definition function $M=\frac1f$ satisfies the condition (refer to (\ref{f_V1}))
\begin{gather}\la{e2}x_aM_a=-2M.\end{gather}
On the other hand,  function $g=g({\bf x})$ in (\ref{Q1}) should depend  on the Euler angles and be independent on the radial variable otherwise the dilatation transformation will change such integral of motion. It follows from the above that function $N=gM$ satisfies the same additional condition  as $M$, i.e.,
\begin{gather}\la{uu}x_aN_a=-2N.\end{gather}

Multiplying (\ref{e1}) by $x_a$, summing up with respect to the repeating index $a$ and using (\ref{e2}) we can express the unknown function $N$ via the bilinear combination variables  $x_a$ and derivatives $M_b$:
\begin{gather}\la{e3}N=\lambda^{ab}x_aM_b\end{gather}

Up to rotation transformations which belong to the equivalence group symmetric tensor $\lambda^{ab}$ can be reduced to the diagonal  form. Moreover, we can restrict ourselves to the following nonzero entries
\begin{gather}\la{e4}\lambda^{11}=\mu, \  \lambda^{33}=\nu\end{gather}
since it is possible to include   nonzero $\lambda^{22}$ to $g( {\bf x})$ by adding the term $P_a\lambda^{22}P_a$ to $Q$.

The corresponding  equation (\ref{e1})  is qualitatively different in two cases:  $\mu\nu=0$ and $\mu\nu\neq0$.
In the first case  we have only one nonzero coefficient, say, $\nu$. Setting $\nu=1$ we come to
 the following version of equation (\ref{e1}):
\begin{gather}\la{e5}M_3+N_3=0,\ N_1=N_2=0.\end{gather}

 The system of equations (\ref{e2}), (\ref{e5}) is solved by the following functions
\begin{gather}\la{sol1}M=\frac{F(\varphi)}{\tilde r^2}-\frac{c_1}{x_3^2},\ N=\frac{c_1}{x_3^2}\end{gather}
where $F(\varphi)$ is an arbitrary function, $\tilde r^2=x_1^2+x_2^2$ and $c_1$ is the integration constant.

Solutions of equations (\ref{me2}) are analogous to (\ref{sol1}) but include another arbitrary function and integration constant.
Substituting these solutions into formulae (\ref{MN}) we come to the inverse mass and potential presented in the following formulae
\begin{gather}\la{T2.1}f=\frac{x_3^2\tilde r^2}{ F(\varphi)x_3^2-c_1\tilde r^2} , \ V=\frac{G(\varphi)x_3^2
 +c_2 \tilde r^2}{  F(\varphi)x_3^2-c_1\tilde r^2}\end{gather}
 while the related integral of motion has the following form:
\begin{gather}\la{Q4}Q=P_3^2-\left(\frac{c_1  }{x_3^2}\cdot H\right)+\frac{c_2}{ x_3^2}.\end{gather}
where we use the notation
\begin{gather}\la{DEF}(N\cdot H)=\sqrt{N}H\sqrt{N}\end{gather} (in our case $N=\frac{c_1}{x_3^2}$).

One more integral of motion can be obtained from (\ref{Q4}) using  symmetry transformation (\ref{IT}). It looks as follows:
\begin{gather}\la{Q5}\tilde Q=K_3^2-\left(\frac{c_1 r^4 }{x_3^2}\cdot H\right)+\frac{c_2 r^4}{ x_3^2}
+3r^2+15x_3^2\end{gather}

In the case when both coefficients $\mu$ and $\nu$
are nontrivial equation  (\ref{e1}) is reduced to the following system
\begin{gather}\la{e51}\begin{split}&\mu M_1+N_1=0,\\& \nu M_3+N_3=0,\\& N_2=0\end{split}\end{gather}
which is solved by the following functions
\begin{gather}\la{e52}M=\frac{c_3}{x_1^2}+\frac{c_4}{x_2^2}+\frac{c_5}{x_3^2}, \ N=\frac{c_3c_1}{x_1^2}+\frac{c_2c_4}{x_3^2}.\end{gather}

Comparing (\ref{e52}) with (\ref{sol1}) it is not difficult to understand that in fact we have two symmetries corresponding to $c_1=0, c_3\neq 0$ and $c_1\neq0, c_3= 0$ and deal with their linear combination. Moreover, since all variables $x_1, x_2$ and $x_3$ are involved into $M$ presented in (\ref{e52}) in the completely symmetric form we can predict one more symmetry whose differential part is equal to $P_2^2$.

Substituting solutions (\ref{e52}) into formulae (\ref{MN}) and  (\ref{MN!}) we obtain the following forms of the inverse mass and potential:
\begin{gather}\la{T2.4}\begin{split}&f=\frac{x_1^2 x_2^2 x_3^2}{c_1 x_2^2x_3^2+c_2x_1^2x_3^2+c_3x_1^2
x_2^2},\\& V=\frac{c_4 x_2^2x_3^2+c_5x_1^2x_3^2+c_6x_1^2
x_2^2}{c_1 x_2^2x_3^2+c_2x_1^2x_3^2+c_3x_1^2
x_2^2}\end{split}\end{gather}
The corresponding integrals of motion are similar to (\ref{Q4}) and have the following form:
\begin{gather}\la{T2.4Q}\begin{split}&Q_1=P_1^2-\left(\frac{c_1}{x_1^2}\cdot H\right)+\frac{c_4}{x_1^2}, \\& Q_2=P_2^2-\left(\frac{c_2}{x_2^2}\cdot H\right)+\frac{c_5}{x_2^2},\\&Q_3=P_3^2-\left(\frac{c_3}{x_3^2}\cdot H\right)+\frac{c_6}{x_1^2}.\end{split}\end{gather}

Notice that only two integrals (\ref{T2.4Q}) are linearly independent since the condition $Q_1+Q_2+Q_3=0$ is satisfied. Two more linearly independent integrals of motion can be obtained by applying symmetry transformation   (\ref{IT}) to $Q_1$ and $Q_2$.

\subsection{Symmetries with differential terms $Q_{(n)}$ (\ref{Q0})  linear in spatial variables}
Thus we find the generic form of PDM Hamiltonians admitting integrals of motion whose differential terms do not depend on spatial variables. The next step is to classify the systems whose integrals of motion are generated by the Killing tensors (\ref{mm1}) which are linear in these variables.

Substituting (\ref{mm1}) into (\ref{mek}) we come to the following equation
\begin{gather}\la{e6}R^a\equiv A^{ab}M_b+N_a=0\end{gather}
where
\begin{gather}\la{e66}A^{ab}=(\mu^{ab}-\lambda^ax^b).\end{gather}
We see that thanks to the presence of the additional condition generated by the dilatation symmetry
equation (\ref{mek}) can be transformed to the linear and homogeneous algebraic constraint (\ref{e66}) for the first order derivatives $N_a$ an $M_b$.
\subsubsection{Admissible combinations of arbitrary parameters}
Derivatives $N_a$ present in equation  (\ref{e66}) can be excluded.
Moreover, it can be done in two independent ways. First, we can calculate the curl of  vector $R^a$ and obtain the following second order system:
\begin{gather}\la{e67}(A^{ab}M_b)_c=(A^{cb}M_b)_a.\end{gather}

One more way to exclude $N_a$ is to use the condition
\begin{gather}\la{e7}x_aN_a=-N\end{gather}
which should be satisfied to guarantee that the dilatation transformation does not destroy the intagral of motion.   Indeed, multiplying   equation (\ref{e6}) by $x_a$, summing up with respect to the repeating index $a$ and using  (\ref{e2}),  (\ref{e7}) we obtain the following expression for $N$:
\begin{gather}\la{e8}N=r^2\lambda^bM_b+\varepsilon^{bcd}\lambda^{cn}x_nx_dM_b.\end{gather}
Substituting (\ref{e8}) into (\ref{e6}) we come to the following system for $M$:
\begin{gather}\la{e68}A^{ab}M_b+2x_a(\lambda^bM_b)+
r^2\lambda^bM_{ab}+\varepsilon^{bcd}\lambda^{cn}(x_nx_dM_b)_a=0.\end{gather}

Equations  (\ref{e2}) , (\ref{e6}), (\ref{e8}) look rather gentle but in fact it is a rather complicated system including eight arbitrary parameters. Fortunately  the number of these parameters can be reduced to five using the equivalence transformations which include rotations and dilatations. Indeed, like in previous section we can apply the rotations to reduce  tensor components $\lambda^{ab}$ to the form presented in (\ref{e4}).

The next reduction of the number of arbitrary parameters can be obtained using the following speculations. System  (\ref{e67}),  (\ref{e68}) includes six linear equations for function $M$ dependent on two variables, refer to (\ref{f_V1}). Substituting the latter expression for $M$ into (\ref{e67}) and  (\ref{e68}) we come to the system of six linear and homogeneous algebraic relations for  variables $X_1=M_\varphi, X_2=M_\theta, X_3=M_{\varphi \varphi}, X_4=M_{\varphi \theta}, X_5=M_{\theta\theta}$ and $X_6M$ where $M_\varphi=\frac{\p M}{ \p \varphi}$, etc., which has the following form:
\begin{gather}\la{e69}B^{\mu\nu}X_\nu=0, \mu, \nu=1,2,...,6,\end{gather}
for the exact expressions of the matrix elements $B^{\mu\nu}$.

Equation (\ref{e69})  represents a necessary condition for the solvability of  system  (\ref{e2}) , (\ref{e6}) and (\ref{e8}). Thus to have a chance to obtain a nontrivial solution of the latter system we are supposed to choose such values of arbitrary parameters  $\lambda_1, \lambda_2, \lambda_3$ and $\lambda_{11}, \lambda_{11}$  which correspond to zero values of the determinant of the matrix whose entries are $B^{\mu\nu}$. The admissible sets of such parameters are given in the following formulae:
\begin{gather}\la{e11a}\mu=0, \ \nu=0, \ \lambda_a {\text{ are  arbitrary }}, \\
\lambda_1 = 0, \ \lambda_2 = 0, \mu=0,  \lambda_3=\pm \nu\la{e11b}\\
 \la{e11g}  \mu=-\nu, \lambda_1 = 0, \ \lambda_2 = \pm\nu,  \ \lambda_3=0\\
 \la{e11c}
\lambda_1 = \lambda_2=\lambda_3 = 0, \mu, \nu   {\text{ are  arbitrary }}.
\end{gather}

In other words effectively we have four versions of system (\ref{e8}) any of which include as maximum two arbitrary parameters since up to rotation transformations in the case (\ref{e11a}) we can restrict ourselves to the only nonzero parameter $\lambda^3$.

\subsubsection{Exact solutions for the arbitrary elements}

Let us consider versions (\ref{e11a})-(\ref{e11g}) step by step. In the case  (\ref{e11a}) we can choose the only nonzero parameter be  $\lambda^3$ and equation (\ref{e6}) is reduced to the following system:
\begin{gather}\la{e12}x_aM_3+N_a=0, a=1,2,3.\end{gather}
In addition, unknowns $M$ and $N$ should satisfy the  conditions (\ref{e2}) and (\ref{e7}): which we rewrite once more for the readers convenience:
\begin{gather}\la{e13}\begin{split}&x_aM_a+2M=0, \\&x_aN_a+N=0.\end{split}\end{gather}
System (\ref{e12}), (\ref{e13}) is easy solvable, its solutions   have the following form:
\begin{gather}\la{T2.2}M=\frac{F(\varphi)r+c_1x_3}{\tilde r^2r}, \ N=\frac{c_1}r.
\end{gather}
while the corresponding integrals of motion look as:
\begin{gather}\la{T2.2Q}\begin{split}&Q_1=\{P_3,D\}-\frac{c_2}{r}+\left(\frac{c_1}{r}\cdot H\right),\\&Q_2=\{K_3,D\}+( r \cdot H)-c_2r-15x_3\end{split}\end{gather}
and are connected via the symmetry transformation (\ref{IT}).

Considering version (\ref{e11b}) we come to the following form of equations (\ref{e6}):
\begin{gather}\la{e14}\begin{split}&\nu(\pm x_1-x_2)M_3+N_1=0,\\&\nu(x_1\pm x_2)M_3+N_2=0,\end{split}\\\la{e15}
\nu(x_1M_2-x_2M_1)\pm x_3M_3+N_3=0.\end{gather}
The evident algebraic consequences of  the subsystem   (\ref{e14}) look as follows
\begin{gather}\la{e16}\begin{split}&\nu\tilde r^2M_3+x_1N_1+x_2N_2=0,\\
&\pm\nu \tilde r^2M_3+x_1N_2-x_2N_1=0.\end{split}\end{gather}

In cylindric variables equations (\ref{e16}) and  (\ref{e15}) take especially simple form:
\begin{gather}\la{e17}\begin{split}&\nu\tilde r^2M_3+\tilde r N_{\tilde r}=0,\\&\pm\nu\tilde r^2M_3+N_\varphi=0,\\&\nu M_\varphi\pm x_3M_3+N_3=0 \end{split}\end{gather}
where $M_{\tilde r}=\frac{\p M}{\p \tilde r}$, etc..

 System (\ref{e17}) is completely equivalent to (\ref{e14}), (\ref{e15}). In addition, we have equations (\ref{e13}) whose form in the cylindric variables is:
 \begin{gather}\la{e18}\begin{split}&\tilde rM_{\tilde r}+x_3M_3+2M=0,\\&\tilde rN_{\tilde r}+x_3N_3+N=0.\end{split}\end{gather}

  The generic solutions of the overdetermined system (\ref{e17}), (\ref{e18})  have the following form:
\begin{gather}\la{e19}\begin{split}&M=\frac{c_1(x_3^2+r^2)e^{\pm2\varphi}+
c_2x_3\tilde re^{\pm\varphi}+c_3\tilde r^2}{\tilde r^4}, \\&
 N=\nu\left( \frac{c_2e^{\pm\varphi}}{\tilde r}+\frac{c_1x_3e^{\pm2\varphi}}{\tilde r^2}\right).\end{split}\end{gather}

The next step is to consider version (\ref{e11g}). We choose the subversion $\lambda^2=+\nu$ which generates the following  form of equations (\ref{e6}):
\begin{gather}\la{e20}\begin{split}&\nu((x_1+x_3)M_3-2x_2M_1)+N_1=0,\\&\nu(x_3M_1+x_2M_2+x_1M_3)+N_2=0,
\\&\nu(x_1+x_3)M_2-2x_2M_1+N_3=0.\end{split}\end{gather}
System  (\ref{e20}) is solved by the following functions:
\begin{gather}\la{e21}\begin{split}&M=\frac{c_1}{y_+^2}+\frac{c_2}{y_-^2}+\frac{c_3x_2}
{y_+^2\sqrt{y_1^2+2x_2^2}},\\& N=\frac{c_3 \sqrt{y_+^2+2x_2^2}}{y_+^2}+x_2\left(\frac{c_1}{y_+^2}
+\frac{c_2}{y_-^2}\right)\end{split}\end{gather}
where $y_\pm=x_1\pm x_3$.

Notice that the subversion $\lambda^2=-\nu$ generates the analogous solutions for $M$ and $N$ where, however, $y_+\to y_-$ and $y_-\to y_+$.

The next note is that applying the rotation around the second coordinate axis we can transform the pair $y_+, y_-$ to $x_1$ and $x_3$. As a result we come to the following realization:
\begin{gather}\la{e21a}\begin{split}&f=\frac{c_1}{x_1^2}+\frac{c_2}{x_3^2}+\frac{c_3x_2}{\tilde r x_1^2}, \\&
 N=\frac{c_1x_2}{x_1^2}+\frac{c_3(x_1^2+2x_2^2)}{2\tilde r x_1^2}.\end{split}\end{gather}

The remaining version (\ref{e11c}) corresponds to the following form of equations  (\ref{e6}):
\begin{gather}\la{e22}\begin{split}&x_2(\mu-\nu)M_3-\mu x_3M_2+N_1=0;\\&\nu x_1M_3-\mu x_3M_1+N_2=0,\\&x_2(\mu-\nu)M_1+\nu x_1M_2+N_3=0.\end{split}\end{gather}
This system added by equations (\ref{e13}) has only trivial solutions provided $\mu$ and $\nu$ are arbitrary. For some special values of $\mu$ and $\nu$ there are nontrivial solutions which, however, are not interesting. Namely for $\mu=\nu, \mu=-\nu , \mu=0$  and $\mu =2\nu$ equations (\ref{e22}) are solved by
\begin{gather*}M=\frac{c_1}{x_2^2+ x_3^2},
\\M=\frac{c_1(x_1^2+x_3^2)+c_2x_1x_3}{(x_1^2-x_3^2)^2},\\
M=\frac{c_1}{x_1^2+ x_2^2}
\end{gather*}
and
\begin{gather*}M=\frac{c_1(x_1^2+x_2^2)+c_2x_1x_2}{(x_1^2-x_2^2)^2}\end{gather*}
correspondingly.

Notice that any of the above presented functions $M$ depend on two variables. The related PDM systems are scale invariant and, in addition,  are invariant w.r.t. shifts along some fixed lines and so admit two parametric Lie groups.
Such systems are completely classified in \cite{AG1} and we will not discuss them here.

Summarising, there are three inequivalent PDM systems which are scale invariant and admit second order integrals of motion generated by the Killing tensors linear in independent variables. The related arbitrary elements are given by formulae (\ref{T2.2}), (\ref{e19}) and (\ref{e21}). In addition, there are such systems admitting two parametric Lie groups which are enumerated in \cite{AG1}.

\subsection{Symmetries with differential terms $Q_{(n)}$ (\ref{Q1}) quadratic in independent variables}
The last class of symmetries which we are supposed to find for the scale invariant systems are integrals of
motion (\ref{Q}) generated by Killing tensors (\ref{mm2}). The specificity  of this class is that it includes
integrals of motions with different transformation properties with respect to the discrete symmetry (\ref{IT}).
Namely, the vector integrals of motion generated by $\mu_{(1)}^{ab}$ are invariant with respect to
transformations (\ref{IT}) while the tensor integrals of motion generated by $\mu_{(2)}^{ab}$  change their sign
under this transformation. It means that in fact we have two independent subclasses of integrals of motion
since the linear combinations of integrals of motion with different transformations properties with respect to
symmetry transformations are forbidden.
\subsubsection{Vector integrals of motion}
Let us start with  vector integrals of motion which are generated by the Killing tensor $\mu_{(1)}^{ab}$
 (\ref{mm2}).  Up to rotation transformations the only nonzero parameter in
$\mu_{(1)}^{ab}$is $\lambda^3$, and the related equations (\ref{e6}) are reduced to the following form:
\begin{gather}\la{e22}\begin{split}&(x_1^2-x_2^2)M_2-2x_1x_2M_1-x_3x_2M_3+N_1=0,\\
&(x_1^2-x_2^2)M_1+2x_1x_2M_2+x_3x_1M_3+N_2=0,\\&x_3x_1M_2-x_3x_2M_1+N_3=0.\end{split}\end{gather}

In addition we have to take into account condition (\ref{e2}) while the analogue of equations (\ref{uu}) and  (\ref{e7}) takes the following form:
\begin{gather}\la{con1}x_aN_a=c_1.\end{gather}
Indeed, tensor $\mu_{(1)}^{ab}$ satisfies the condition
$x_c\mu_{(1)c}^{ab}=2\mu_{(1)}^{ab}$, thus the related operator $Q_{(n)}$ (\ref{Q1}) is not changed under the
dilatation transformation. The same property can be requested to the total integral of motion (\ref{Q0})
presented in (\ref{Q1}). In other words, we can suppose that the first term $Q_{(0)}=P_agP_a$ in (\ref{Q1}) is not changed also. It is the case when function $g=\frac{N}M$ satisfies the same condition as  $\mu_{(1)}^{ab}$, i.e., $x_ag_a=2g$.  It is the case if $N$ satisfies condition (\ref{con1}) with $c_1=0$ since $M$ satisfies (\ref{e2}).

But what has happen in the case when parameter $c_1$ in (\ref{con1}) is not trivial? The related integral of motion will be changed under the conformal transformations, but this change is reduced to the form $Q\to Q+c H$ where $H$ is the Hamiltonian and $c$ is a constant. Such changes are surely admissible since $H$ commutes with itself.

The presented speculations are not necessary since equations  (\ref{e22}) can be solved directly without using (\ref{con1}). However, the latter condition presents a nice tool to simplify calculations.

Multiplying the first of equations (\ref{e22}) by $x_1$, the second equation by $x_2$, the third  by $x_3$, summing up the obtained results and using (\ref{con1}) we obtain the following consequence:
\begin{gather}r^2(x_1M_2-x_2M_1)=-c_1.\la{e23}\end{gather}
Its generic solution is
\begin{gather}\la{e23}M=-\frac{c_1\varphi+G(\theta)}{r^2}\end{gather}
where $\varphi$ and $\theta$ are the Euler angles and $G(\theta)$ is an arbitrary function. This solution is valid for the initial equations   (\ref{e22}) also.
\subsubsection{Tensor integrals of motion}

Consider now the tensor integrals of motion generated by $\mu_{(2)}^{ab}$ (\ref{mm2}). Up to rotation transformations  we can restrict ourselves to the following version of nonzero  parameters $\lambda^{ab} $:
\begin{gather}\la{e18}\lambda^{33} =1\end{gather}
or
\begin{gather}\la{e19a}\lambda^{11} =\mu, \ \lambda^{22} =\nu\end{gather}
where $\mu$ and $\nu$ are arbitrary real constants.

In the case (\ref{e18})  integral of motion (\ref{Q}) and determining equations (\ref{e6}) take the following forms:
\begin{gather}\la{e20a}Q=L_3^2+\left(N\cdot H\right)-\tilde N\end{gather}
and
\begin{gather}\la{e20}\begin{split}&
x_3x_\alpha M_3=N_\alpha, \ \ \ \alpha=1,2,\\&
\tilde r^2M_3-x_3x_aM_a=N_3.\end{split}\end{gather}
The system (\ref{e13}), (\ref{e20}) is solved by the following functions
\begin{gather}\la{e25}M=\frac{G(\theta)+F(\varphi)}{\tilde r^2}, \ N=F(\varphi)\end{gather}
where $G(\theta)$ and $F(\varphi)$ are arbitrary functions of Euler angles.

The case (\ref{e19a}) is the most complicated and interesting. The related integral of motion (\ref{Q}) takes the form
\begin{gather}\la{e19b}Q=\mu L_1^2+\nu L_2^2+Q_{(0)} +\eta\end{gather}
and determining equations (\ref{e6}) are reduced to the following system:
\begin{gather}\la{e22}\begin{split}&\mu(x_2^2+x_3^2)M_1-\nu x_1x_2M_2+N_1=0,\\&\nu(x_1^2+x_3^2)M_2-\mu x_1x_2M_1+N_2\end{split}\end{gather}
where we use relations (\ref{e13}) and (\ref{con1}) to simplify the system.

Without loss of generality parameters $\mu$ and $\nu$ are supposed to be non equal, i.e., $\mu\neq \nu$, since for $\mu=\nu$ operator (\ref{e19b}) is proportional to a linear combination of the squared dilatation generator $D^2$ and operator (\ref{e20a}) analysed in the above.

Excluding unknown variable $N$ we come to the following second order equation for $M$:
\begin{gather}\la{e23}\left(\mu(x_2^2+x_3^2)-\nu(x_1^2+x_3^2)\right)M_{12}+x_1x_2(\mu M_{11}-\nu M_{22})+3(\mu x_2M_1-\nu x_1M_2)=0\end{gather}
which looks as drastically complicated one. However, changing dependent  and independent variables in the following way:
\begin{gather*}M=\frac{F(\omega,z)}{\omega x_3^2}\end{gather*}
where
\begin{gather}\la{ou}\begin{split}&\omega=\sqrt{z^2+4\rho}, \
\rho=\frac{\mu x_1^2-\nu x_2^2}{x_3^2}+\mu\nu(\mu-\nu),\\&
 z=\frac{\nu x_1^2+\mu x_2^2+
 (\mu-\nu)(\mu^2-\nu^2)x_3^2}{x_3^2\sqrt{|\mu\nu(\mu-\nu)|}}\end{split}\end{gather}
 we can reduce it to the standard D'Alembert equation for function $F(\omega,z)$
 \begin{gather*}(\p_{\omega\omega}-\p_{zz})F(\omega,z)=0\end{gather*}
 and so
 \begin{gather}\la{M}M=\frac{F(\omega+z)+G(\omega-z)}{ x_3^2}.\end{gather}

 For any fixed $F(\omega+z)$ and $G(\omega-z)$ we can solve the system  (\ref{e22}), (\ref{con1})  and find functions $N = N(F,G)$
corresponding to $M$ defined in (\ref{M}). Unfortunately, it is seemed be impossible to represent
functions $N = N(F,G)$ in closed form for $F$ and $G$ arbitrary. However, it is possible do it  for
 rather extended classes of functions $F$ and $G$. In particular it can be done  if the sum $F(.) + G(.)$ is a
homogeneous function of $z$ and $\omega$, say, if $F(.) + G(.)=\Phi_m$
where
\begin{gather}\la{e24} \Phi_m=\frac{(\omega+z)^m+(-1)^{m+1}(\omega-z)^m}{ \omega}\end{gather}

Substituting (\ref{e24}) into (\ref{M}) and integrating the corresponding equations (\ref{e22}), (\ref{con1})  we find the related
functions $M$ and $N$ in closed form for $m$ arbitrary:
\begin{gather}\la{e26}M=\frac{\Phi_m}{ x_3^2}, \ N=-\rho\Phi_{m-1}.\end{gather}

Notice that linear combinations of solutions (\ref{e25}) also solve equations (\ref{e22}) and (\ref{con1}).

We find all inequivalent solutions of the determining equations  (\ref{mek}). Thus  we  fix all inequivalent PDM
systems which are scale invariant and admit at least one second order integral of motion.
By definition such systems are integrable. However, some of them admit more than one integral of  motion as
it is shown in the following subsection.

\subsubsection{Search for superintegrable sustems}

We find all inequivalent solutions of the determining equations  (\ref{mek}). Thus  we  fix all inequivalent PDM
systems which are scale invariant and admit at least one second order integral of motion.
By definition such systems are integrable.

The found mass functions $M$ include arbitrary coefficients and even arbitrary functions.
 For some  special form of these arbitrary functions and special combinations of the values of arbitrary parameters the related PDM system can admit more integrals of motion and be superintrgrable or even maximally superintegrable. Just these possibilities are studied in the present section.

 Let us consider step by step all solutions presented in formulae (\ref{sol1}), (\ref{T2.4}), (\ref{T2.2}), (\ref{e19}), (\ref{e21a}), (\ref{e25}), (\ref{M}) and search for the cases when they satisfy the following functional equations:
 \begin{gather}\la{e27}M=M'\end{gather}
 where $M$ is the solution given by one of the mentioned formulae and $M'$ is a solution given by another one.

Let us start with the case presented in (\ref{T2.4}). In this case we have three integrals of motion fixed in
 (\ref{T2.4Q}), two of which are functionally independent.  One more integral of motion is the generator of
  conformal transformations $D$. Thus the related PDM system is  superintegrable.

Comparing solutions (\ref{T2.4}) with (\ref{e25}) we observe that they coincide provided functions $F(\varphi)$
and $G(\theta)$ present in (\ref{e25}) have the following forms:
\begin{gather}\la{WAU}F(\varphi)=\frac{c_1x_2^2}{x_1^2}+\frac{c_2x_1^2}{x_2^2}-c_1-c_2, \ G(\theta)=
\frac{c_3\tilde r^2}{x_3^2}.\end{gather}
The related function $N$ is equal to $F(\varphi)$ given in (\ref{WAU}) and integral of motion  (\ref{e20a}) is reduced
to the following form:
\begin{gather}\la{WAU1}Q=L_3^2-\left(\frac{c_1x_2^4+c_2x_1^4}{x_1^2x_2^2}\cdot H\right)+
\frac{c_4x_2^4+c_5x_1^4}{x_1^2x_2^2}.\end{gather}

Since variables $x_1, x_2$ and $x_3$ are involved into formulae  (\ref{T2.4}) in completely symmetric way we
can conclude that there exist two more integrals of motion admitted by the considered system which
 can be obtained  from (\ref{WAU1}) by the changes $L_3\to L_1, x_3\to x_1, x_2\to  x_2,  x_1\to -x_3$
 and $L_3\to L_2, x_1\to x_1, x_2\to x_3, x_3\to -x_2$.

 Let us compare solutions for $M$ given by relations (\ref{sol1}), (\ref{T2.2}), (\ref{e21a}) and (\ref{e25}). We recognize that functions (\ref{sol1}) and  (\ref{T2.2}) are particular cases of the function presented in (\ref{e25}) with
 $G(\theta)=\tan(\theta)^2 $ and $G(\theta)=\cot(\theta) $ respectively. The same is true for solution (\ref{e21a}) which has the form (\ref{e25}) (and (\ref{sol1})!) with $G(\theta)=c_3\tan(\theta)^2$ and $F(\varphi)=\frac{c_1}{\cos(\varphi)^2}+\frac{c_2\tan(\varphi)}{\cos(\theta)}$.  On the other hand functions  (\ref{sol1}) and  (\ref{T2.2}) are essentially different and cannot be reduced one to another for some particular form of arbitrary functions or  nontrivial arbitrary constants. The case of the trivial constants $c_1$ is forbidden since in this case we have as minimum two parametric Lie symmetry group.

 It follows from the above that the PDM systems whose masses have the form given by equations (\ref{T2.2}), (\ref{e21a}) and (\ref{e25}) admit the additional integral of motion presented in (\ref{Q4}) and so are superintegrable. Moreover, the mass function (\ref{e21a}) is maximally superintegrable since in addition it admits one more integral of motion whose form is presented in (\ref{e20a}).

The   mass functions  presented in   (\ref{e19}) and (\ref{e23}) have rather specific dependence on the angular variables which has nothing to do with the other found versions of such functions. Any of the  related PDN systems admits  only one integral of motion  additional to $D$ and is integrable but not superintegrable.

The only version which was not considered jet is given by equation  (\ref{M}). It is extremely difficult to look
directly for the cases when it can coincide with one of the version presented in
(\ref{sol1}), (\ref{T2.4}), (\ref{T2.2}), (\ref{e19}), (\ref{e21a}), (\ref{e25}), (\ref{M}). However we can search for
the cases when equations (\ref{e23}) are compatible with equations (\ref{e67}) corresponding to these versions.
As a result we find the only case when it is possible, and this case corresponds to the following particular
form of functions  (\ref{M}) and (\ref{e25}) which coincide:
\begin{gather*}M=\frac{c_1}{x_1^2}+\frac{c_2}{x_2^2}+\frac{c_3}{x_3^2}+\frac{c_4}{r^2}\end{gather*}
while the related potential (\ref{MN!}) looks as follows:
\begin{gather*}V=\frac{r^2(c_5x_2^2x_3^2+c_6x_1^2x_3^2+c_7x_1^2x_2^2)
+c_8x_1^2x_2^2x_3^2}
{r^2(c_1x_2^2x_3^2+c_2x_1^2x_3^2+c_3x_1^2x_2^2)+c_4x_1^2x_2^2x_3^2}\end{gather*}
The corresponding integrals of motion are given by equation (\ref{e20a}) and
where $N=\frac{c_1}{\cos(\varphi)^2}+\frac{c_2}{\sin(\varphi)^2}$ and
$\tilde N=\frac{c_5}{\cos(\varphi)^2}+\frac{c_6}{\sin(\varphi)^2}.$
\subsubsection{Classification results}
Thus we find all inequivalent position dependent masses which correspond to
 systems which are scale invariant and admit at least one second order integral of motion.

In this section we summarize the obtained results.  They are represented in Table 1
where $F(.), G(.), \tilde F(.), \tilde G(.)$ are arbitrary functions whose arguments are fixed in the brackets,
$\omega, z, \rho$ and $\Phi_m$ are given by relations  (\ref{ou}) and (\ref{e24}),
$\hat \eta^a=3r^2+15x_a^2, a=1, 2, 3$, $W=x_1^2x_2^2x_3^2,$
$c_1, c_2, ...$ are arbitrary real parameters. In addition, notation (\ref{DEF}) is used.

\newpage
{\begin{center}Table 1.  Inverse   masses, potentials  and  integrals of motion for  scale invariant
systems.\end{center}}
\small{
\begin{tabular}{c c c c}
\hline
No&$f$&$V$&\text{Integrals of motion}\\
\hline
1\vspace{1.0mm}&$\frac{r^2\tilde r^2}{r^2F(\theta)+c_1\tilde r^2\varphi}$&$\frac{r^2G(\theta)+c_2\tilde r^2\varphi}{r^2F(\theta)+c_1\tilde r^2\varphi}$&$\begin{array}{c}\{L_3,D\}-(c_1\ln(r)\cdot H)\\+c_2\ln(r)\end{array}$\\

\hline

2\vspace{0.5mm}&$ \frac{\tilde r^2}{G(\theta)+F(\varphi)}$&$\frac{\tilde G(\theta)+
\tilde F(\varphi)}{G(\theta)+F(\varphi)}$&$\begin{array}{c}
L_3^2- (F(\varphi)\cdot H)+\tilde F(\varphi)\end{array}$\\

\hline

3\vspace{0.5mm}&$ \frac{\tilde r^4}{c_1(x_3^2+r^2)e^{\pm2\varphi}+c_2x_3\tilde re^{\pm\varphi}+c_3\tilde r^2}$&$\frac{c_4(x_3^2+r^2)e^{\pm2\varphi}+c_5x_3\tilde re^{\pm\varphi}+c_6\tilde r^2}{c_1(x_3^2+r^2)e^{\pm2\varphi}+c_2x_3\tilde re^{\pm\varphi}+c_3\tilde r^2}$&$\begin{array}{c}\{P_3,(D\mp L_3)\}\\+2\left((\frac{c_2e^{\pm\varphi}}{\tilde r}+\frac{c_1x_3e^{\pm2\varphi}}{\tilde r^2})\cdot H\right)\\-2(\frac{c_5e^{\pm\varphi}}{\tilde r}+\frac{c_4x_3e^{\pm2\varphi}}{\tilde r^2}),\\\{K_3,(D\mp L_3)\}-15x_3\\+2\left(r^2(\frac{c_2e^{\pm\varphi}}{\tilde r}+\frac{c_1x_3e^{\pm2\varphi}}{\tilde r^2})\cdot H\right)\\-2r^2(\frac{c_5e^{\pm\varphi}}{\tilde r}+\frac{4_6x_3e^{\pm2\varphi}}{\tilde r^2})
\end{array}$\\
\hline
4\vspace{0.5mm}&$\frac{x_3^2\omega}{F({\omega+z})+
G(\omega-z)}
$&
$\frac{\tilde F({\omega+z})+
\tilde G(\omega-z)}{F({\omega+z})+
G(\omega-z)}
$&$\begin{array}{c}\mu\{P_1,K_1\}+\nu\{P_2,K_2\}\\+\left(g(F,G)\cdot H\right)+\eta(\tilde F,\tilde G)
\end{array}$\\

\hline

5\vspace{0.5mm}&$\begin{array}{c}\sum_m c_m\Phi_m,\end{array}$&$\begin{array}{c}\frac{\sum_m \tilde c_m\Phi_m}{\sum_m c_m\Phi_m}\end{array}$&$
\begin{array}{c}\mu\{P_1,K_1\}+\kappa\{P_2,K_2\}\\-\left(\rho\sum_m  c_m\Phi_{m-1}\cdot H\right)\\+\rho\sum_m \tilde c_m \Phi_{m-1}-\hat\eta
\end{array}$\\

\hline

6\vspace{0.5mm}&$\frac{x_3^2\tilde r^2}{ c_1\tilde r^2+F(\varphi)x_3^2} $&$\frac{G(\varphi)x_3^2
 +c_2 \tilde r^2}{ c_1\tilde r^2+F(\varphi)x_3^2}$&
$\begin{array}{c}P_3^2+\frac{c_2}{ x_3^2}-\left(\frac{c_1  }{x_3^2}\cdot H\right),\\
K_3^2+\frac{c_2 r^4}{ x_3^2}-\left(\frac{c_1 r^4 }{x_3^2}\cdot H\right)+\hat \eta^3,\\L_3^2+G(\varphi)- (F(\varphi)\cdot
H)\end{array}$\\

\hline

7\vspace{0.5mm}&$ \frac{\tilde r^2r}{F(\varphi)r+c_1x_3}$&$\frac{G(\varphi)+c_2x_3}
{F(\varphi)r+c_1x_3}$&$\begin{array}{c}
\{P_3,D\}-\frac{c_2}{r}+(\frac{c_1}{r}\cdot H),\\\{K_3,D\}+( r \cdot H)-c_2r-15x_3,\\L_3^2+G(\varphi)-\left(F(\varphi)\cdot H\right)\end{array}$\\

\hline

8\vspace{0.5mm}&$\frac{r^2W}{r^2(c_1 x_2^2x_3^2+c_2x_1^2x_3^2+c_3x_1^2
x_2^2)+c_4W}$&$\frac{r^2(c_5 x_2^2x_3^2+c_6x_1^2x_3^2+c_7x_1^2
x_2^2)+c_8W}{r^2(c_1 x_2^2x_3^2+c_2x_1^2x_3^2+c_3x_1^2
x_2^2)+c_4W}$&$\begin{array}{c}
L_1^2-\left(\frac{c_3x_2^4+c_2x_3^4}{x_2^2x_3^2}\cdot H\right)+\frac{c_6x_2^4+c_5x_3^4}{x_2^2x_3^2},\\

L_2^2-\left(\frac{c_3x_1^4+c_1x_3^4}{x_1^2x_3^2}\cdot H\right)+\frac{c_6x_1^4+c_3x_3^4}{x_1^2x_3^2}\end{array}$\\

\hline

9\vspace{1mm}&$\frac{x_3^2x_1^2\tilde r}{c_1 \tilde r x_3^2+c_2 x_2x_3^2+c_3 x_1^2\tilde
r}$&$\frac{c_4 \tilde r x_3^2+c_5 x_2x_3^2+c_6 x_1^2\tilde
r}{c_1 \tilde r x_3^2+c_2 x_2x_3^2+c_3 x_1^2\tilde
r}$&$\begin{array}{c} \{L_3,P_1\}+\frac{2c_4 x_2}{x_1^2}+\frac{c_5(\tilde
r^2+x_2^2)}{\tilde r x_1^2}\\-\left(H\cdot(\frac{2c_1 x_2}{x_1^2}+\frac{c_2(\tilde
r^2+x_2^2)}{\tilde r x_1^2})\right),\\\{L_3,K_1\}+\frac{2c_4 x_2r^2}{x_1^2}+\frac{c_5r^2(\tilde
r^2+x_2^2)}{\tilde r x_1^2}\\-\left(H\cdot(\frac{2 r^2c_1
x_2}{x_1^2}+\frac{c_2 r^2(\tilde r^2+x_2^2)}{\tilde r x_1^2})\right),\\L_3^2-\left(\frac{c_1\tilde
r^2+c_2 x_2 \tilde r}{x_1^2}\cdot H\right)+\frac{c_4\tilde
r^2+c_5 x_2 \tilde r}{x_1^2},\\

P_3^2-\left(\frac{c_3  }{x_3^2}\cdot H\right)+\frac{c_6}{ x_3^2},\\

K_3^2-\left(\frac{c_3 r^4 }{x_3^2}\cdot H\right)+\frac{c_6 r^4}{ x_3^2}+\hat \eta^3

\end{array}$\\

\hline

10\vspace{0.5mm}&$\frac{x_1^2 x_2^2 x_3^2}{c_1 x_2^2x_3^2+c_2x_1^2x_3^2+c_3x_1^2
x_2^2}$&$\frac{c_4 x_2^2x_3^2+c_5x_1^2x_3^2+c_6x_1^2
x_2^2}{c_1 x_2^2x_3^2+c_2x_1^2x_3^2+c_3x_1^2
x_2^2}$&$\begin{array}{c} P_2^2-\left(\frac{c_2}{x_2^2}\cdot H\right)+\frac{c_5}{x_2^2},\\
P_1^2-\left(\frac{c_1}{x_1^2}\cdot H\right)+\frac{c_4}{x_1^2},\\

K_2^2-\left(\frac{c_2r^4}{x_2^2}\cdot H\right)+\frac{c_5r^4}{x_2^2}+\hat \eta^2,\\
K_1^2-\left(\frac{c_1r^4}{x_1^2}\cdot H\right)+\frac{c_4r^4}{x_1^2}+\hat \eta^1,\\

L_1^2-\left(\frac{c_3x_2^4+c_2x_3^4}{x_2^2x_3^2}\cdot H\right)+\frac{c_6x_2^4+c_5x_3^4}{x_2^2x_3^2},\\

L_2^2-\left(\frac{c_3x_1^4+c_1x_3^4}{x_1^2x_3^2}\cdot H\right)+\frac{c_6x_1^4+c_3x_3^4}{x_1^2x_3^2}
\end{array}$\\
\hline\hline

\end{tabular}}

Notice that the systems whose arbitrary elements are present items 1-5 are integrable
while while the remaining items represent  superintegrable systems. Moreover, the systems fixed in Items
9 and 10
are maximally superintegrable.

A bit mysterious system  presented in Item 10 admits as much as seven linearly
independent  integrals of motion, namely,
\begin{gather}\la{LA1}\begin{split}&Q_1=P_1^2-\left(\frac{c_1}{x_1^2}\cdot H\right)+\frac{c_4}{x_1^2},\\
&Q_2=P_2^2-\left(\frac{c_2}{x_2^2}\cdot H\right)+\frac{c_5}{x_2^2},\\
&Q_3=L_1^2-\left(\frac{c_3x_2^4+c_2x_3^4}{x_2^2x_3^2}\cdot H\right)+\frac{c_6x_2^4+c_5x_3^4}{x_2^2x_3^2},
\\&Q_4=
L_2^2-\left(\frac{c_3x_1^4+c_1x_3^4}{x_1^2x_3^2}\cdot H\right)+\frac{c_6x_1^4+c_3x_3^4}{x_1^2x_3^2},\\
&Q_5=K_1^2-\left(\frac{c_1r^4}{x_1^2}\cdot H\right)+\frac{c_4r^4}{x_1^2}+\hat \eta^1,\\
&Q_6=K_2^2-\left(\frac{c_2r^4}{x_2^2}\cdot H\right)+\frac{c_5r^4}{x_2^2}+\hat \eta^2,\\&Q_7=D.
\end{split}\end{gather}

Operators $Q_5$ and $Q_6$ are connected  with $Q_1$  and $Q_2$ via discrete transformation (\ref{IT}).
Since the  maximal number of integrals of motion   allowed for Hamiltonian systems with three degrees
 of freedom is equal to 4, the remaining integrals of motion $Q_1-Q_4$ and $Q_5$ should
 be functionally dependent. This dependence is implicit and can be found by calculation  the double commutator
 $[Q_1-Q_2,[Q_1-Q_2,Q_3+Q_4]]$.

\section{PDM systems invariant with respect to shifts}
The last class of PDM system  we discus in the present paper are those ones which by definition admit Lie symmetry whose generator is $P_3$.  Since the related arbitrary elements
$M$ and $V$ do not depend on $x_3$ it is possible a priori reduce the number of admissible second
order integrals of motion.

Let Q be an integral of motion (\ref{Q}) admitted by equation (1) with arbitrary elements (\ref{f_V2}).
By definition $P_3$ is the  integral of motion too, the same is true for the commutators
$[P_3, Q], [P_3, [P_3, Q]]$ and $[P_3, [P_3, [P_3, Q]]].$ Thus any second order symmetry induces the symmetry
generated by Killing tensors $\mu^{ab}$  (\ref{K1})-(\ref{K9}) independent on $x_3$ which are itemized below together with one additional special tensor:
\begin{itemize}\item $\mu_1^{ab}$ with arbitrary $\lambda_1^{ab}$,
\item  $\mu_2^{ab}$  with $\lambda_2^{3}=0$, \item $\mu_3^{ab}$ and $\mu_6^{ab}$ where the only nonzero parameters are $\lambda_3^{33}$ and $\lambda_6^{33},$
 \item $\mu_2^{ab}$ with $\lambda_2^\alpha=0, \ \lambda_2^3\neq0$
    \end{itemize} refer to (\ref{K1})-(\ref{K9}).

    The  integrals of motion (\ref{Q}) generated by $\mu^{ab}$ specified in the first three items commute with $P_3$ and so do not induce additional symmetries. However, the analogous property belongs to integral of motion generated by $\mu_2^{ab}$
with $\lambda_2^3\neq 0$. Its commutator with $P_3$ is not trivial but proportional to $P_3^2$ which is by now means an  additional symmetry for the considered class of PDM systems.

    Classifying PDM systems admitting these induced symmetries we can obtain the complete list of the
    systems invariant with respect to shifts and admitting second order integrals of motion.
    Then it will be necessary to specify the additional integrals of motion which can be admitted by the found
    systems.
    As a result we obtain the complete list of the systems presented  in the following Tables 2-4.
    The detailed calculations are presented in Sections 4.1 - 4.3.

\subsection{Integrals of motion whose differential part $Q_{(n)}$ (\ref{Q1}) is independent on
$x_3$ }

The first step of our programm is to classify the systems whose integrals of motion are generated by the
Killing tensors specified at the beginning of Section 4, i.e., by the following linear combinations:
\begin{gather}\la{eq6} \begin{split}&\mu^{\alpha\beta}=\lambda^{\alpha\beta}_1+ \lambda_2^\alpha x_\beta+ \lambda_2^\beta x_\alpha+\lambda_6^{33}(\delta^{\alpha\beta}\tilde r^2-x_\alpha x_\beta),\\&\mu^{3\alpha}=\lambda^{33}_3\varepsilon^{3\alpha c}x_c+\lambda_1^{3\alpha}+\lambda_2^3x_\alpha\end{split}\end{gather} where $\alpha$ and $\beta$ can take the values 1, 2.

Using the rotations around the third coordinate axis we  reduce $\lambda_2^2$ to zero, and denote  $\lambda_6^{33}=\sigma, \lambda_1^{11}=-\lambda_1^{22}=\omega, \lambda_1^{12}=\lambda_1^{21}=\kappa, \lambda^3_2=\alpha, \lambda_3^{33}=\mu$.
The corresponding integrals of motion (\ref{Q}) take the following form:
\begin{gather}\la{0}Q=Q^{(1)}+Q^{(2)}\end{gather}
where
\begin{gather}\la{eq7a}Q^{(1)}=\omega P_1^2+\kappa P_1P_2+\nu \{P_1, L_3\}+\sigma L_3^2
+(N^{(1)}\cdot H)+\eta^{(1)},\\\la{eq7a1}
Q^{(2)}=\lambda^{3\alpha} P_3P_\alpha+\mu L_3P_3+\alpha\{P_3,D\}+(N^{(2)}\cdot H)+\eta^{(2)}.\end{gather}
The related determining equations (\ref{mek}) are reduced to the following system:
\begin{gather}\la{eq7}\begin{split}&(x_2(\sigma x_2-2\nu)+\omega)M_1
+((x_1(\nu-\sigma x_2)+\kappa)M_2+N_1=0,\\&(x_1(\nu-\sigma x_2)+\kappa)M_1+(\sigma x_1^2
-\omega)M_2+N_2=0,\end{split}\\\la{eq7aa}(\lambda^{31}-\mu x_2+
\alpha x_1)M_1+(\mu x_1+\lambda^{32}+\alpha x_2)M_2+2\alpha M+N_3=0.\end{gather}

Equations (\ref{eq7}), (\ref{eq7aa}) include eight arbitrary parameters. But in fact their number can be essentially
reduced using shift transformations and analysing this system consistency. Just such reduction is the subject
of the following subsection.

The specific property of system   (\ref{eq7}), (\ref{eq7aa}) is that that equations (\ref{eq7}) include the set of arbitrary
parameters $\sigma, \nu, \omega, \kappa$  and do not include $\alpha, \mu, \lambda^{31}, \lambda^{32}$
while equation  (\ref{eq7aa}) includes the second set of parameters and do not include the first one.
It means that system  (\ref{eq7}), (\ref{eq7aa}) with $\sigma=\nu=\omega=\kappa =0$ represents
  the determining equations for integral
of motion   $Q^{(2)}$. Moreover, in the case when $\alpha, \mu, \lambda^{31}$ and $\lambda^{32}$ are trivial
but some of the other coefficients are nonzero we have the  determining equations for  $Q^{(1)}$ . In the another
combinations of the arbitrary parameters the discussed system defines the PDM Schr\"odinger equations admissing
both integrals of motion, i.e., $Q^{(1)}$ and $Q^{(2)}$.

\subsubsection{Systems admitting integral of motion $Q^{(2)}$}

Let us start with the case when parameters $\sigma, \nu, \omega$ and $ \kappa$ are trivial. The
corresponding subsystem  (\ref{eq7}) is reduced to the conditions $N_1=N_2=0$ so the related function $N$ can depend
on $x_3$ only. Moreover, since the l.h.s. of equations (\ref{eq7aa}), does not depend on $x_3$, the generic form of function
$N$ looks as follows:
\begin{gather}\la{eq7b}N=2\lambda x_3 +G(x_1,x_2)\end{gather}
where $\lambda$ is a constant.

Substituting (\ref{eq7b}) into (\ref{eq7aa}) we come to the following equation:
\begin{gather}\la{eq7c}(\lambda^{31}-\mu x_2+
\alpha x_1)M_1+(\mu x_1+\lambda^{32}+\alpha x_2)M_2+2\alpha M+2\lambda=0.\end{gather}

Let at least one of parameters $\mu$ or $\alpha$ be nonzero. Then, using shifts of variables $x_1$ and $x_2$ we can reduce  to zero
$\lambda^{31}$ and $\lambda^{32}$. After that we  can integrate equation (\ref{eq7c}) and obtain
the generic form of the corresponding function $M$ :
\begin{gather}\la{s1}M=\frac{F(\alpha \ln(\tilde r)-\mu\varphi)}{\tilde r^2}-\frac\lambda\alpha, \ \alpha\lambda\neq0\end{gather}
\begin{gather}\la{s2}M=F(\tilde r)-2\lambda \varphi,\ \alpha=0, \ \mu\neq0,\end{gather}
and
\begin{gather}\la{s3}M=\frac{F(\varphi)}{\tilde r^2}-\frac{\lambda}{\alpha} , \alpha\neq0, \ \mu=0.\end{gather}

For both $\alpha$ and $ \mu $ trivial
 we  cannot a priori reduce reduce $\lambda^{31}$ and $\lambda^{32} $ to zero, but can normalize them by
 simultaneous scaling of all independent variables $x_1, x_2$ and $ x_3$. The corresponding equation
 (\ref{eq7c}) again can be solved exactly. The related  generic form of $M$ is:
 \begin{gather}\la{s4}M=-\lambda y_1+F(y_2)\end{gather}
 where $y_1=\lambda^{31}x_1+\lambda^{32}x_2, \ y_2=\lambda^{31}x_2-\lambda^{32}x_1 $
 and the related parameters $\alpha, \mu, \lambda^{31}$ and $\lambda^{32}$ are supposed to satisfy the
 following condition:
 \begin{gather}\la{s5}\alpha =0, \ \mu=0, (\lambda^{31})^2+(\lambda^{31})^2=1.\end{gather}
Using rotations around the third coordinate axis we can reduce $y_1$ to $x_1$ and $y_2$ to zero.

Thus we
 fix four inequivalent versions of parameters  $\alpha, \mu, \lambda^{31}$ and $\lambda^{32}$  and find the generic
 form of the corresponding mass functions $M$ which
 are presented in equations (\ref{s1}), (\ref{s2}) (\ref{s3}) and (\ref{s5}). The corresponding potentials  are easy
 calculated using definition (\ref{MN!}).

  The related PDM system admit second order integrals of motion
 presented in (\ref{eq7a1}) where  $N=c_1x_3$. In other words, these integrals of motion have the following forms:
 \begin{gather}\la{s23} Q^{(2)}=\alpha\{P_3,D\}+\mu L_3P_3+2\lambda(x_3,\cdot H)+\lambda^{31}P_3P_1-2\tilde\lambda x_3\end{gather}
 where parameters $\alpha$ and $\mu$ take the values fixed in    (\ref{s1}) - (\ref{s3}).

\subsubsection{Systems admitting integral of motion $Q^{(1)}$}

  One more possibility corresponds to
  the case when all
 parameters $\alpha, \mu, \lambda^{31}$ and $\lambda^{32}$ are trivial, and system   (\ref{eq7}),
 (\ref{eq7aa}) is reduced to its subsystem  (\ref{eq7}) added by the condition $N_3=0.$ As in the previous subsection
 we have  four arbitrary parameters only. Moreover, their number of this parameters can be reduced.
 Indeed, let  parameter $\sigma$ is nontrivial making the  shift $x_2\to x_2+\frac\nu\sigma $ we can nullify
 parameter $\nu$. If $\sigma$ is trivial but $\nu$ is nonzero we can nullify parameter $\omega$ by
 the shift  $x_2\to x_2-\frac{\omega}{2\nu}$.

 In all the cases discussed above it is also  possible to nullify parameter $\kappa$ making the
 shift
 $x_1\to x_1-\frac{\kappa}{\nu}.$
 Whenever both $\sigma$ and $\nu $ we also can nullify $\kappa$ making the rotation around
 the third coordinate axis.

 In other words, there are the following inequivalent versions of parameters $\sigma, \nu$ and $\omega$ which
 have to be considered:
 \begin{gather}\la{s8}\alpha=\mu=\lambda^{31}=\lambda^{32}=\kappa=0.\end{gather}
Moreover, conditions  (\ref{s8}) are added by one of the following ones:
 \begin{gather}\la{s6}\sigma=0, \nu\neq0, \ \omega=0,\\\la{s77}\sigma\neq 0, \nu=0, \ \omega=0,\\
 \la{s7777}
 \sigma= 0, \nu=0, \ \omega\neq0\end{gather}

 The next step is to find inequivalent solutions of equations  (\ref{eq7}), (\ref{eq7aa}) for all versions of arbitrary
 parameters enumerated in the above.

The related equation  (\ref{eq7aa}) is reduced to the condition $N_3=0$
and there are four versions of equations  (\ref{eq7}) corresponding to the conditions (\ref{s6}), (\ref{s77}),
and   (\ref{s7777}) .

 Let us  differentiate the first
of equations (\ref{eq7}) with respect to $x_2$ and the second of these equations with respect to $x_1$ and equate
the obtained expressions. As a result we come to the differential consequence of  (\ref{eq7})   which does not include unknown $N$:
\begin{gather}\la{s9}\begin{split}&\left(\sigma(x_2^2-x_1^2)-2\nu x_2+\omega\right)M_{1,2}
+(\sigma x_1x_2-\nu x_1 -\kappa)(M_{1,1}-M_{2,2})\\&+3(\sigma x_2-\nu)M_1
-3\sigma x_1M_2=0.\end{split}\end{gather}

If conditions  (\ref{s8}) and (\ref{s6}) are valid equations  (\ref{s9}) and (\ref{eq7}) reduced to the following forms:
\begin{gather}\la{s24}x_1(M_{2,2}-M_{1,1})-2x_2M_{1,2}-3M_1=0\end{gather}
and
\begin{gather}\la{s25}\begin{split}&2x_2M_1-x_1M_2-N_1=0,\\&
x_1M_1+N_2=0.\end{split}\end{gather}

It is not too easy task to find the generic solution of the above presented equations. However, changing
the dependent and independent variables
$(x_1, x_2)\to(z_1, z_2), \ M(x_1, x_2)\to {\Phi(z_+, z_-)}$ where
\begin{gather}\la{s26}\Phi(z_+, z_-)=\tilde r M(x_1, x_2), \ z_\pm=\tilde r\pm x_2\end{gather}
it is possible to reduce (\ref{s24}) to the d'Alembert equation $\Phi_{z_+,z_-}$ for $\Phi=\Phi(z_+, z_-)$. Thus
the generic form of solution for equation (\ref{s24}) is
\begin{gather}\la{s26} M=\frac1{\tilde r}(F(\tilde r+x_2)+G(\tilde r-x_2))\end{gather}
where $F(\tilde r+x_2)$ and $G(\tilde r-x_2)$ are arbitrary functions whose arguments are
fixed in the brackets. The corresponding potential (\ref{MN!}) looks as follows:
\begin{gather}\la{s27}V=\frac{W(z_{+})+K(z_{-})}{ F(z_{+})+G(z_{-})}.\end{gather}

Solving equations  (\ref{s25}) where $M$ is function (\ref{s26}) we obtain the related function $N$ in the form
\begin{gather}\la{s27}N={x_2}M+G(z_-)-F(z_+).\end{gather}

In accordance with the above the PDM system with the mass and potential given by relations (\ref{s26}) and
 (\ref{s27}) admits the integral of motion of the following form:
\begin{gather}\la{s29}Q=\{L_3,P_1\}+P_ax_2P_a-((F(z_{+})-G(z_{-}))\cdot H)+W(z_{+})-K(z_{-}).\end{gather}

The next version of the arbitrary parameters which we consider is given by relations (\ref{s8}) and (\ref{s77}).
The corresponding equation (\ref{s9}) is reduced to the following one:
\begin{gather}\la{s29q}(x_2^2-x_1^2)M_{1,2}
+ x_1x_2(M_{1,1}-M_{2,2})+3( x_2M_1
- x_1M_2)=0\end{gather}
which is solved by the following function:
\begin{gather}\la{s30}M=F(\tilde r)-\frac {G(\varphi)}{\tilde r^2}\end{gather}
while the related potential $V$ (\ref{MN!}) and function $N$ satisfying equations (\ref{eq7}) have the following form:
\begin{gather}\la{s31}V=\frac{\tilde F(\tilde r)-\tilde G(\varphi)}{F(\tilde r)-G(\varphi)}, \ N=G(\varphi).
\end{gather}

The PDM system whose mass and potential are given by relations (\ref{s30}) and (\ref{s31}) admits the following
integral of motion:
\begin{gather}\la{s32}Q=L_3^2+(G(\varphi)\cdot H)-\tilde G(\varphi).\end{gather}

The next version (\ref{s6}), (\ref{s77}) corresponds to the most complicated determining equations.
Up to scaling of variables $x_a$ we can set $\sigma=\pm1$ and reduce (\ref{eq7a}) to the following form:
\begin{gather}\la{eq161}\begin{split}&( x_2^2-1)M_1- x_1 x_2M_2+N_1=0,\\& (x_1^2-1) M_2-x_1 x_2M_1+N_2=0,\ N_3=0\end{split}\end{gather}
while the corresponding differential consequence (\ref{s9}) looks as follows:
\begin{gather}\la{eq17a}(x_2^2-x_1^2-2\sigma)M_{12}+x_1x_2(M_{11}-M_{22})+3(x_2F_1-x_1F_2)=0\end{gather}
an is rather cumbersome.  However, choosing new dependent and independent  variables, namely,
\begin{gather}\la{eq18}\hat M= \omega M, \
\tilde x_1=\omega+\tilde r^2,\ \tilde x_2=\omega-\tilde r^2 \end{gather}
where $\omega=\sqrt{\tilde r^4-4y}, y=x_1^2-x_2^2-\sigma$ we can reduce (\ref{s9})
to the standard D'Alembert equation for $\hat M=\hat M(\tilde x_1,\tilde x_2)$ whose solutions are
\begin{gather}\la{eq19}\hat M=F_1(\tilde x_1)+F_2(\tilde x_2) \end{gather}
where $F_1(.)$ and $F_2(.)$ are arbitrary functions. Thus the generic solution for the mass function is:
\begin{gather}\la{eq18n}M=\frac{F_1(\tilde x_1)+F_2(\tilde x_2)}\omega.\end{gather}

For any fixed functions $F_1(.)$ and $F_2(.)$  we can solve equation (\ref{eq161}) and find the corresponding function $N$. Unfortunately it is impossible to find $N$ in closed form for $F_1(.)$ and $F_2(.)$  arbitrary.
However, it can be done for the special class of these functions. Namely, considering the polynomial functions
\begin{gather}\la{eq19aa}F_1(\tilde x_1)+F_2(\tilde x_2)=c_m\Omega_m\end{gather}
where
\begin{gather}\la{eq191}\Omega_m={(\omega+\tilde r^2)^m+
(-1)^{m+1}(\omega-\tilde r^2)^m}\end{gather} and summation is imposed over the repeating indices
over all natural values $m=1,2,..$ .

Substituting the corresponding expression for $M$, i.e. $M=\frac{c_m\Omega_m}\omega$ into equations
 (\ref{eq161})
we come to the relatively simple solvable system whose solutions are
\begin{gather}\la{eqqq}N=\frac{-2yc_m\Omega_{m-1}}\omega.\end{gather}

At this point we stop our discussion of the most complicated version (\ref{s6}), (\ref{s77})
Notice that the corresponding integrals of motion (\ref{eq7a}) is reduced to the following form
\begin{gather}\la{s40}Q=L_3^2+(P_1^2-P_2^2)+\left(N(F,G)\cdot H\right)-\tilde N(\tilde F,\tilde G)\end{gather}
were $N(F,G)$ and $\tilde N(F,G)$ are functions specified in the above.
\

The last version we are supposed to consider is presented by relations (\ref{s7777}). The corresponding equation
  (\ref{s9}) is reduced to the standard  D'Alembert form $M_{1,2}=0$ and is solved by the following
  function:
  \begin{gather}\la{s31a}M=F(x_1)+G(x_2).\end{gather}
 The related potential $V$ (\ref{MN!}) and function $N$ satisfying condition (\ref{eq7}) have the following form:
\begin{gather}V=\frac{\tilde F(x_1)+\tilde G(x_2)}{F(x_1)+G(x_2)}, \ N_{(1)}=G(x_2)-F(x_1)\end{gather}
and the corresponding integral of motion (\ref{eq7a}) is reduced to the following form:
\begin{gather}\la{s31b}Q=P_1^2 +\left((G(x_2)-F(x_1))\cdot H\right)+\tilde F_1(x_1)-\tilde G(x_2).\end{gather}
Our search for shift invariant PDM systems admitting a second order integral of motion in form
 (\ref{eq7a1}) turned to the end. The classification results obtained in this and previous sections are
 summarized in Table 2.

 \subsection{Search for superintegrable systems}

 In previous subsection we classified integrable system which admit at least one second order integral of motion.
 The found masses and potentials are defined op to arbitrary functions.

 For some special forms of these functions the system can obtain more integrals of motion and be superintegrable or
 even maximally superintegrable. Just just systems are classified in the present section.

 There exist three versions of the considered superintegrable systems. Namely, they can admit two integrals of motion
 of type (\ref{eq7a}), two integrals of motion
 of type (\ref{eq7a}), or  integrals of motion of the both types. All these versions are studied in the following subsections.

 \subsubsection{Syperintegrable systems admitting integrals of motion of type (\ref{eq7a})}

 The PDM  systems admitting one  integral of motion of  type (\ref{eq7a}) are classified in section 3.2.1 where
 four inequivalent systems having this property are presented. The related integrals of motion are given by
 equations (\ref{s29}),   (\ref{s32}), (\ref{s40}) and (\ref{s31b}). We are supposed to consider the cases when
 these integrals of motion are admitted together with one more integral of motion of generic form (\ref{eq7a}).
 Fortunately, the cases  when such defined pairs of integrals of motion include  (\ref{s32})  or (\ref{s40})
 can be omitted as being equivalent to the remaining cases up to shift and rotation transformation, and it is
 sufficient to restrict ourselves to the cases (\ref{s29}) and (\ref{s31b}).

 Let us start with the system whose mass depends on two arbitrary functions an is presented by equation
 (\ref{s31a}).   Such system admits integral of motion (\ref{s31b}). Our task is to search the cases when such system
 admits one more integral of motion of the generic form  (\ref{eq7a}) which can happen for some special forms
 of the  mentioned functions.

 Substituting   (\ref{s32}) into equation (\ref{s9}) were without loss of generality we can set $\omega=0$ we obtain
 the following relation:
\begin{gather}\la{s33}((\sigma x_2-\nu)x_1-\kappa)(F_{1,1}-G_{2,2}+3(\sigma x_2-\nu)F_1-3\sigma x_1G_2=0.
\end{gather}

Let $\sigma\neq 0$ then we can set $\sigma=1$ and nullify $\nu$ making the shift $x_2\to x_2+{\nu}$.
As a result we come to equation (\ref{s33}) which is solved by the following functions:
\begin{gather*} F(x_1)=c_1x_1^2+c_2, \ G(x_2)=c_1x_2^2+c_3.\end{gather*}
The corresponding function (\ref{s31a}) has the additional Lie symmetry with respect to rotations around
the third coordinate axis and so can be ignored.

If in addition $\kappa=0$ equation (\ref{s33}) is solved by functions $F(x_1)=c_1x_1^2+c_2 x_1^{-2},
\ G(x_2)=c_1x_2^2+c_3 x_2^{-2}+c_4$. The related functions $M$  (\ref{s31a}) and $N$ look as follows:
\begin{gather}\la{s34}M=c_1\tilde r^2+\frac{c_2}{x_1^2}+\frac{c_3}{x_2^2}+c_4,
\ N=-\frac{c_2 x_2^2}{x_1^2}-\frac{c_3 x_1^2}{x_2^2}\end{gather}
and integral of motion (\ref{eq7a}) is reduced to the following form:
\begin{gather}\la{s34a} Q=L_3^2-(N\cdot H)+\frac{c_6 x_2^2}{x_1^2}+\frac{c_7 x_1^2}{x_2^2}.\end{gather}

If parameter $\sigma$ in (\ref{s33}) is trivial but $\nu\neq0$  we set $\nu=1$ and nullify $\kappa$ making the shift
 $x_1\to x_1-{\kappa}$. Solutions of the such reduced equation (\ref{s33}) generate the following function $M$
 of generic form (\ref{s31a}), potential $V$ (\ref{MN!}) and function $N$ solving equations (\ref{eq7}):
\begin{gather}\la{s35}\begin{array}{l}M=\frac{c_1}{x_1^2}+c_2(x_1^2+4x_2^2)+c_3x_2+c_4,\\
V=\frac{c_1+c_2(x_1^2+4x_2^2)+x_1^2(c_3x_2+4_8)}{c_5+c_6(x_1^2+4x_2^2)+x_1^2(c_7x_2+c_8)},
 \ \ N=\frac1{2x_1^2}(4x_2(c_1-c_2x_1^4)-c_3x_1^4).\end{array}\end{gather}

 Finally, if both $\sigma$ and $\nu$ are trivial, equations (\ref{s33}) and (\ref{eq7}) are  easy solvable and
 generate the following $M, V$ and $N$:
 \begin{gather}\la{s35a}\begin{array}{l}M=c_1x_1+c_2x_2+c_3\tilde r^2+c_4, \\ N=-(2c_3x_1x_2+c_2x_1+c_1x_2),
 \\ V= \frac{c_5x_1+c_6x_2+c_7\tilde r^2+c_8}{c_1x_1+c_2x_2+c_3\tilde r^2+c_4}.\end{array}\end{gather}

 The PDM systems whose mass and potential are fixed by relations (\ref{s35}) and  (\ref{s35a}) admit the
 second integrals of motion in form (\ref{eq7a}) where functions $\tilde N$ has the same form as $N$
 where, however, parameters $c_k$ are changed to $c_{k+4}$, and values of parameters
 $\sigma, \nu, \kappa$ are specified in the above.

 The next step is to consider the system admitting integral of motion (\ref{s32}). The corresponding mass
 (\ref{s30}) satisfies equation (\ref{s29q}) and depends on two arbitrary function. Out task is to find such special
 form of these functions corresponding
 to the case  when such system admits one more integral of motion of the generic form   (\ref{eq7a}). Substituting
 It happens in the case when the system of equations (\ref{s24}) and (\ref{s8}) is consistent.

Substituting  (\ref{s30})  into (\ref{s8}) we come to the following equation for functions $F(\tilde r) $ and $G(\varphi)$:
\begin{gather}\la{EQ}\begin{array}{l}\left(\kappa\cos(2\varphi)-\frac{\omega}2\sin(2\varphi)\right)(\tilde r^5 F_{\tilde r,\tilde r}-
\tilde r^4F_{\tilde r}+2\tilde r G_{\varphi,\varphi}-8G_{\varphi})\\+
\nu\cos(\varphi)(\tilde r^6F_{\tilde r,\tilde r}+4\tilde r^5F_{\tilde r}+\tilde r^2G_{\varphi,\varphi}-2G)
+3\sin(\varphi)\tilde r^2G_{\varphi}=0\end{array}\end{gather} where by definition $\nu\neq 0$ since for
$\nu$ zero we come to the case rotationally equivalent to (\ref{s35}).

Equation (\ref{EQ}) is consistent for arbitrary values of parameters $\nu, \kappa$ and $\omega$. However,
if $\kappa$ is not trivial, its solutions are $F(r)=c_1+\frac{c_2}{\tilde r^2}$ and $G(\varphi)=c_2$ , and the
related mass (\ref{s30}) is  constant.

Let both $\kappa$ and $\omega$ be trivial then equation (\ref{EQ}) is solved by the following function:
\begin{gather}\la{s36}F(\tilde r)=c_1+\frac{c_2}{\tilde r}+\frac{c_5}{\tilde r^2}, \
 G(\varphi)=c_5-\frac{c_3\sin(\varphi)}{\cos(\varphi)^2}-\frac{c_4}{\cos(\varphi)^2}\end{gather}
while the corresponding mass (\ref{s30}),  potential (\ref{MN!}) and function $N$ solving equations (\ref{eq7}) are:
\begin{gather}\la{s37}\begin{split}&M=c_1+\frac{c_2}{\tilde r}+\frac{c_3}{x_1^2}+\frac{c_4x_2}{\tilde r x_1^2},
\\ &V= \frac{c_6x_1^2+c_8x_2+c_5\tilde r x_1^2+c_7\tilde r}{c_2x_1^2+c_4x_2+c_1\tilde r x_1^2+c_3\tilde r}.
\end{split}\end{gather}
and
\begin{gather}\la{s38}N=x_2\left(\frac{c_2}{\tilde r}+\frac{c_3}{x_1^2}\right)+
\frac{c_4(x_1^2+2x_2^2)}{\tilde r x_1^2}.\end{gather}

The related PDM system admits two second order integrals of motion, namely, operator (\ref{s32}) with $G(\varphi)$
given by relation (\ref{s36}) where $c_5=0$ and operator given by relation  (\ref{eq7a}) where the only
nonzero parameter is $\nu=1$ and $N $ is function (\ref{s38}).

If $\kappa=0$ but both $\nu$ and $\omega$ are nontrivial, equation (\ref{EQ}) is solved by functions (\ref{s36})
with $c_2=c_4=0$. The corresponding PDM system admits two the above mentioned integrals of motion with
trivial $c_2$ and $c_3\neq0$ and one more integral of motion $Q= P_1^2-\left(\frac{c_3}{x_1^2}\cdot H\right)
+\frac{c_6}{x_1^2}$. The obtained results are presented in Item 3 of Table 4.

 \subsubsection{Superintegrable systems admitting integrals of motion of type (\ref{eq7a1})}

Consider now superintegrable systems admitting two integrals of motion of generic form (\ref{eq7a1}). The rules
of game are analogous to ones used in the previous subsection. Namely, to classify such systems it is necessary to
consider step by step all inequivalent versions of mass functions presented in section 4.1 and test their
compatibility with generic equation (\ref{eq7c}).

Let us start with the mass function presented in (\ref{s4}). Substituting  it into  (\ref{eq7c}) where
$\lambda\to\tilde\lambda$ we reduce the
 latter equation to the following form:
 \begin{gather}\la{s41}(\alpha x_2+\mu x_1+\lambda^{32})F_2+2\alpha F-\lambda \mu x_2+
 3\alpha \lambda x_1+\tilde \lambda=0\end{gather}
 where $F=F(x_2)$.

 Equation (\ref{s41}) has consistent solutions for $F$ in the unique   case, namely, when
 $\mu=\alpha=0, \lambda^{32}\neq0$. Choosing such values of the arbitrary coefficients and setting
 $\lambda^{32}=1$ we come to
  the following solution:
   \begin{gather}\la{s42}F=-\tilde\lambda x_2+c_1\end{gather}
   which generates the following function (\ref{s4}) and potential (\ref{MN!}):
   \begin{gather}\la{s43}\begin{split}&M=\lambda x_1-\tilde \lambda x_2+c_1,\\&
   V=\frac{c_1x_1+c_3x_2+c_3}{\lambda x_1-\tilde \lambda x_2+c_1}.\end{split}\end{gather}

 The PDM system with the mass  and potential presented in (\ref{s43}) admits  two second order integrals
 of motion of type  (\ref{eq7a1}), namely
 \begin{gather}\la{s44} Q_1=\{P_3,P_1\}+\lambda x_3\end{gather}
 and
 \begin{gather}\la{s45} Q_2=\{P_3,P_2\}+\tilde\lambda x_3.\end{gather}
 Moreover, it admits additional integrals of motion as it will be shown in the next subsection.

 \subsubsection{Superintegrable systems admitting integrals of motion of both types}

Let arbitrary parameters satisfy conditions (\ref{s1}). The related function solves equation (\ref{eq7a}) and $M$ is defined up to arbitrary function
which depend on a single variable, and our task is to specify the form of this function and admissible values of arbitrary
parameters which are compatible with equation (\ref{s9}).

In the case when $M$ has the form presented in (\ref{s1}) equation (\ref{s9}) is reduced to the  ordinary
second order equation for function $F=F(z)$ where $z= \alpha \ln(\tilde r)-\mu\varphi$. It has the following form
\begin{gather}\la{s10}\Phi_{(2)} F_{z,z}+\Phi_{(1)} F_z +\Phi_{(0)} F +\Phi=0\end{gather}
where $F_z=\frac{\p F}{\p z}$, etc.,  $\Phi_{(0)}=2\nu x_1
\tilde r^2-8(\omega x_1x_2+\kappa(x_2^2-x_1^2)), \Phi=0, $ $\Phi_{(2)}$ and $\Phi_{(1)}$ are fourth order
polynomials  in $x_1, x_2$.

Since $\Phi_{(0)}$ is not a function of $z$ it follows from (\ref{s10}) that either $\Phi_{(0)}$ or
$F$ should be trivial. To obtain
a nontrivial $F$ we have to nullify $\Phi_{(0)}$ which can be achieved by setting
$\omega=\kappa=\nu=0$. As a result
equation (\ref{s9}) is reduced to the following form:
\begin{gather}\la{s11}2\sigma\mu F_{z,z}=0.\end{gather}

 Since $\mu$ by definition is nontrivial it follows from (\ref{s11}) that there are two possibilities:
 $\sigma=0$, $F(z)$ is an arbitrary function, and
 \begin{gather}\la{s12}F=c_1 z+c_2, \sigma\neq0.\end{gather}

 In the first case we have generic solution (\ref{s1}) for $M$, and the only integral of motion is given by relation
(\ref{eq7a}) where the only nonzero coefficients are $\mu$ and $\alpha$,  i.e.,
  \begin{gather}\la{eq12c}Q^{(1)}=\mu P_3L_3+\alpha\{P_3,D\}
-\left(2c_2\nu x_3\cdot H\right)+2c_5\nu x_3.\end{gather}

In the case (\ref{s12}) there are two integrals of motion, namely, (\ref{eq12c}) and
\begin{gather}\la{eq12d}Q^{(2)}=L_3^2+2(c_1\varphi\cdot H)-2c_4\varphi.\end{gather}
The latter integral of motion is the particular case of (\ref{eq7}) where the only nonzero coefficient is $\sigma$.

The next version we consider is presented in equation(\ref{s2}). Substituting the function $M$ given there into
equation (\ref{s9}) we again come to equation of generic form (\ref{s10}) where $z=\tilde r$ and
\begin{gather}\la{s14}\begin{split}&\Phi_{(2)}=-4\mu \tilde r^4(\nu x_1 \tilde r^2 -\omega x_1x_2+
\kappa(x_1^2-x_2^2)),\ \Phi_{(1)}=6\mu\nu x_1\tilde r^4,
\\&\Phi_{(0)}=0, \ \Phi=2\lambda\sigma\tilde r^4-4x_2\nu \tilde r^2+\omega(x_1^2-x_2^2)+
4\kappa x_1x_2.\end{split}\end{gather}

In accordance with  (\ref{s14}) equation (\ref{s10}) has nontrivial solutions only in the case when all
 parameters $\omega, \nu, \kappa$ and $\sigma$ are equal to zero. Thus solution (\ref{s2}) generates the PDM
 system which admit the only  integral of motion in form  (\ref{eq7a})  where the only nonzero coefficient is $\mu$.
 ,i.e.,
 \begin{gather}\la{s16}Q=\{P_3,L_3\}-c_1(x_3\cdot H)+\lambda x_3. \end{gather}

 Considering version (\ref{s3}) we  come to equation (\ref{s10}) where $z=\varphi$ and
 \begin{gather}\la{s17}\begin{split}&\Phi_{(2)}=(\nu x_1\tilde r^2+\kappa(x_1^2-x_2^2)
 -\omega x_1x_2),\\&\Phi_{(1)}=  -3\nu x_2\tilde r^2-12\kappa x_1x_2+3\omega (x_2^2-x_1^2),\\
 & \Phi=8\omega x_1x_2-2\nu x_1\tilde r^2+8\kappa(x_2^2-x_1^2), \ \Phi_{(0)}=0.\end{split}\end{gather}

 Functions (\ref{s17}) can be reduced to functions of $\varphi$ in two cases, namely, when coefficients
 $\omega, \kappa, \sigma$ and $\nu$ satisfy one of the following relations:
 \begin{gather}\la{s18a}\omega=0, \ \kappa=0, \nu\neq0 \\\la{s18b}  \nu=0.
 \end{gather}

Notice that in the second case we can (and have done it) to nullify also coefficient $\kappa$ applying the rotation
transformation.

 In case (\ref{s18a}) the corresponding equation (\ref{s10}) is solved  by the following function:
 \begin{gather}F(\varphi)=\frac {c_1}{\cos(\varphi)^2}+\frac{c_2\sin(\varphi)}{\cos(\varphi)^2}\la{s19a}\end{gather}
while for the case (\ref{s18b}) we obtain
\begin{gather}\la{s19b}F(\varphi)=\frac {c_1}{\cos(\varphi)^2}+\frac {c_2}{\sin(\varphi)^2}.\end{gather}

The corresponding masses and potentials look as follows:
\begin{gather}\la{s19aa}\begin{split}&M=\frac{c_2}{x_1^2} +\frac{c_1x_2}{\tilde r x_1^2}-{\lambda}, \\
&V=\frac{c_3 \tilde r+c_4x_2+c_5\tilde r x_1^2}{c_1 \tilde r+c_2x_2-\lambda\tilde r x_1^2}\end{split}\end{gather}
for case (\ref{s18a}), and
\begin{gather}\la{s19bb}\begin{split}&M=\frac{c_2}{x_1^2} +\frac{c_2}{ x_1^2}-{\lambda}, \\&
V=\frac{c_3 x_1^2+c_4 x_2^2+c_5x_1^2x_2^2}{c_1 x_1^2+c_23 x_2^2-\lambda x_1^2x_2^2}\end{split}\end{gather}
for the case (\ref{s18b}).

The PDM systems whose masses and potentials are fixed in  (\ref{s19aa}) and (\ref{s19bb}) admit the following
integrals of motion of type (\ref{eq7a})
\begin{gather}\la{s50}Q^{(1)}=\{P_1,L_3\}+
\left(\frac{2c_2\tilde r+c_1(x_1^2+2x_2^2)}{x_1^2\tilde r}\cdot H\right)
-\frac{2c_5\tilde r+c_4(x_1^2+2x_2^2)}{x_1^2\tilde r}\end{gather}
and
\begin{gather}\la{s51}P_1^2-P_2^2+\left(\frac{c_1x_1^2-c_2x_2^2}{x_1^2x_2^2}\cdot H\right)-
\frac{c_4}{x_2^2}+\frac{c_5}{x_1^2}\end{gather}
correspondingly. In addition, both these systems by construction admit integral of motion (\ref{s23}).

\subsubsection{Integrals of motion dependent on $x_3$}

In the previous sections we have found all inequivalent PDM systems which are shift invariant and admit
second order integrals of motion of the generic forms (\ref{eq7a}) and (\ref{eq7a1}).
The specificity of such integrals of motion is that their commutators with symmetry
$P_3$ are trivial or proportional to $P_3$. Moreover, relations (\ref{eq7a}) and (\ref{eq7a1}) represent
the most general forms of second order integrals  of motion with possess this property.
On the other hand, commuting the other possible integrals of motion with $P_3$ $n$ times,
for some $n$  we have to come to a  symmetry of form (\ref{eq7a}) or (\ref{eq7a1}). In other words, we
have found all inequivalent shift invariant systems which admit at least one second order integral of
motion. However, we did not present additional integrals of motion of more general form. Just these
integrals are classified in the previous section.

The first step in the promised classification is to specify  the bilinear combinations of generators
(\ref{QQ}) whose commutators with $P_3$ have the form presented in (\ref{eq7a}) or (\ref{eq7a1}).
To achieve this goal we will use the following commutation relations
\begin{gather}\la{004}\begin{split}&[P_3,P_\alpha]=0, \ [P_3,D]=-\ri P_3,\\
& [P_3,L_1]=\ri P_2, [P_3,L_2]=-\ri P_1, \ [P_3,L_3]=0,\\&
[P_3,K_3]=2\ri D, \ [P_3,K_1]=-2\ri L_2, \  [P_3,K_2]=2\ri L_1.\end{split}\end{gather}

Let us present  the generic form of operators $\tilde Q^{(1)}$  and $\tilde Q^{(2)}$ whose commutator
with $P_3$ is proportional  to $Q^{(1)}$  (\ref{eq7a}) and $Q^{(2)}$ (\ref{eq7a1}) respectively:
\begin{gather}\la{05a}
\tilde Q^{(1)}=\frac\omega2(\{P_1,L_2\}+\{P_2,L_1\})+\kappa(P_1L_1-P_2L_2)  +\nu\{L_2,L_3\}
+\left(N^{(1)}\cdot H\right)+
\tilde N^{(1)},\\\la{05a1}\tilde Q^{(2)} =\mu L_3D+
\frac{\alpha}{2}\{P_3,K_3\}+\frac12(\lambda^{32}\{P_3,L_1\}-\lambda^{31}\{P_3,L_2\}) +\left(N^{(2)}\cdot H\right)+
\tilde N^{(2)}\end{gather}
In accordance with (\ref{004}) operators $Q^{(1)}, Q^{(2)}$ and $\tilde Q^{(1)}, \tilde Q^{(2)}$
satisfy the following  relation
\begin{gather}\la{06a}[P_3,\tilde Q^{(a)}]=\ri Q^{(a)}, a=1, 2\end{gather}
where $Q^{(1)}$ is operator (\ref{eq7a}) and $Q^{(2)}$ is operator (\ref{eq7a1})  with $\sigma=0$. Notice that
equation (\ref{06a}) has no solutions for $a=2, \sigma\neq0$.

Let operator $\tilde Q^{(a)}$ with some combinations of arbitrary parameters
 be an integral of motion of equation (1) with arbitrary elements (\ref{f_V2}). Then operator
 $Q^{(a)}$ with the same values of the mentioned parameters is the integral of motion too.
 Thus the necessary condition of the commutativity $\tilde Q^{(a)}$ with the Hamiltonian $H$ is the
 commutativity of $Q^{(a)}$  with $H$. Since we already found all inequivalent
  operators $Q^{(a)}$, we know all admissible combinations of arbitrary parameters in integrals
  of motion (\ref{05a}). However, the mentioned necessary condition is not sufficient one,
  and so we are supposed to verify whether  operators $\tilde Q^{(a)}$ do commute with the found
  Hamiltonians. Making this routine job  we recover the $x_3$ dependent integrals of motion presented
  in Items  2, 5 of Table 2 and Items 1-4 of Table 3.

A bit more sophisticated speculations are requested to find the integral of motion
$\hat Q=\{P_2,L_1\}+2\left(\frac{c_1x_3}{x_2^2}\cdot H\right)-2\frac{c_2x_3}{x_2^2}$ presented implicitly
in Item 3 of Table 3 as a commutator of the first integral of motion with $P_3$.
It satisfies the  following relation:
\begin{gather}\la{07a}[P_3,\tilde Q]=\ri( Q+P_aP_a-P_3^2)\end{gather}
where $Q$ is a particular case of the integral of motion presented in Item 1 of Table 3:
\begin{gather}\la{08a} Q=P_1^2-P_2^2+\left(\left(\frac{c_1}{x_2^2}-F(x_1)\right)\cdot H\right)-\frac{c_2}{x_2^2}+\tilde F(x_1).\end{gather}

Relation (\ref{07a}) is more general than  (\ref{06a}). Its  right hand size includes two additional terms one of which, i.e., $P_3^2$, by definition commutes with the Hamiltonian, while
 $P_aP_a$ can be included into $Q_0$ by adding the unity to $g({\bf x})$.

 Thus we already know the integrals of motion $\tilde Q$ linear in variable $x_3$. The next step is to find such integrals $\hat Q$ which are quadratic polynomials in this variable. By definition they satisfy the following relation
 \begin{gather}[P_3,\hat Q]=\tilde Q.\end{gather}
 Repeating the speculations presented in the above we find integrals of motion
\begin{gather}\la{L} \begin{split}&\hat Q_1=L_1^2-\left(\frac{c_1x_3^2}{x_2^2}\cdot H\right)+
\frac{c_2x_3^2}{x_2^2}, \\ & \hat Q_2=D^2-c_1(r^2\cdot H)+c_2r^2,\\&\hat Q_3=\{L_2,L_3\}-
\frac{2c_6x_3\tilde r+c_4x_3(x_1^2+2x_2^2)}{x_1^2\tilde r}\\&
+\left(\frac{2c_3x_3\tilde r+c_1x_3(x_1^2+2x_2^2)}{x_1^2\tilde r}\cdot H\right)\end{split}\end{gather}
first of which is presented in Item 3 of Table 3  and Item 2 of Table 4 while the
second and third ones one can be found in Item 6 of Table 3 and  Item 3 of Table 4.

Notice that integrals of motion whose differential part  $Q_{(n)}$ (\ref{Q1})  includes third and second order polynomials in $x_3$ are not admitted by the considered shift invariant PDM systems.
\subsection {Classification tables}

We have completed the classification of the PDM systems which are invariant with respect to rotations around
the third coordinate axis and admit second order integrals of motion. In this section we represent the obtained
results in the three tables which include integrable, superintegrable and maximally superintegrable systems.

In the tables we give the inverse masses and potentials of such systems together with the functionally
independent integrals of motion. Notice that the number of the the linearly independent integrals is more extended.
To find them it is sufficient to calculate the commutators of the presented operators with $P_3$.

In Table 2 we use the following notations:
$\Omega_m={(\tilde\omega+\tilde r^2)^m+(-1)^{m+1}(\tilde\omega-\tilde r^2)^m}, $
$y=x_1^2-x_2^2-1,\ \tilde \omega=\sqrt{\tilde r^4-4y}, c_m$ with $m=1,2,..$ are arbitrary
parameters. In addition, in this and the following system notation (\ref{DEF}) is used.

 Table 2 represents all inequivalent   PDM systems which admit two integrals of motion,
one of which is the generator of shift transformations along the third coordinate axis and the other
is a  second order integral of motion.   In other words the
systems presented in Table 2 are integrable while superintegrable systems are represented in Tables 3
and 4.

The potentials and inverse masses presented in Table 2 include arbitrary functions of some special variables. In other
words the class of such systems is rather extended.

The systems represented in Item 8 are particular cases of the generic system given in Item 7. The masses and
potentials of these
 systems
are polynomials in variables  $\tilde \omega\pm \tilde r^2$ while in Item 7 there are arbitrary functions of these variables.
The nice feature of the polynomial solutions is that the related integrals of motion can be presented explicitly
while for the case of arbitrary functions we are able to represent the constructive elements $N$ and
$\tilde N$ of these integrals of motion only as solutions of the related equations (\ref{eq161}) which can be
solved for any particular case of these arbitrary functions.

\begin{center}Table 2.  Inverse   masses, potentials  and  integrals of motion for integrable shift invariant systems\end{center}
\begin{tabular}{c c c c}
\hline
\vspace{1.5mm}No&$f$&$V$&\text{Integrals of motion}\\
\hline\\
1\vspace{2mm}&$\frac{1}{F(x_1)+G(x_2)}$&$\frac{\tilde F(x_1)+\tilde G(x_2)}{F(x_1)+G(x_2)}$&
$\begin{array}{c}P_1^2-P_2^2+\tilde F(x_1)-\tilde G(x_2)\\+\left((G(x_2)-F(x_1))\cdot H\right)\end{array}$\\

\hline

2\vspace{2mm}&$\frac{\tilde r^2}{F(\tilde r)-G(\varphi)}$&$\frac{\tilde F(\varphi)+
\tilde G(\tilde r)}{G(\varphi)- F(\tilde r)}$&$L_3^2+(G(\varphi)\cdot H)+\tilde G(\varphi)$\\

\hline

3\vspace{2mm}&$\begin{array}{c}\frac{\tilde r}{ F(z_{+})+G(z_{-})},
\\z_{\pm}=\tilde r\pm x_2\end{array}$&$\frac{W(z_{+})+K(z_{-})}{ F(z_{+})+G(z_{-})}$&
$\begin{array}{c}\{L_3,P_1\}+P_ax_2P_a\\-((F(z_{+})-G(z_{-}))\cdot H)\\+W(z_{+})-K(z_{-})\end{array}$\\

\hline

4\vspace{2mm}&$\frac{\tilde r^2}{c_1\tilde r^2+F(\varphi)}$&$\frac{c_2\tilde r^2+
G(\varphi)}{c_1\tilde r^2+F(\varphi)}$&$\{P_3,D\}-2c_1(x_3\cdot H)+2c_2x_3$\\

\hline

5\vspace{2mm}&$\frac{1 }{c_1\varphi+F(\tilde r)}$&$\frac{G(\tilde r)+
c_2\varphi}{c_1\varphi+F(\tilde r)}$&$\{P_3,L_3\}-c_1(x_3\cdot H)+c_2 x_3$\\

\hline

6\vspace{2mm}&$\frac{\tilde r^2 }{c_1 \tilde r^2 +F(\nu\varphi-
\alpha\ln(\tilde r))}$&$\frac{c_2\tilde r^2 +G(\nu \varphi-\alpha\ln(\tilde r))}
{c_1 \tilde r^2 +F(\nu\varphi-
\alpha\ln(\tilde r))}$&$\begin{array}{c}\alpha\{P_3,D\}+\nu\{P_3,L_3\}\\-\left(2c_1 x_3\cdot H\right)+2c_2 x_3\end{array}$\\

\hline

7\vspace{2mm}&$\frac{\tilde \omega}{F(\tilde \omega+\tilde r^2)+G(\tilde\omega-\tilde r^2)}$&
$\frac{\tilde F(\tilde\omega+\tilde r^2)+\tilde G(\tilde\omega-\tilde r^2)}{F(\tilde\omega+\tilde r^2)
+G(\tilde\omega-\tilde r^2)}$&
$\begin{array}{c}L_3^2+(P_1^2-P_2^2)\\+\left(g(F,G)\cdot H\right)-g(\tilde F,\tilde G)\end{array}$
\\

\hline

8\vspace{2mm}&$\begin{array}{c}\frac{\tilde \omega}{\sum_m c_m\Omega_m},\end{array}$&$\begin{array}{c}\frac{\sum_m \tilde c_m\Omega_m}{\sum_m c_m\Omega_m}\end{array}$&$
\begin{array}{c}L_3^2+(P_1^2-P_2^2)\\-2\left(\frac{y\sum_m  c_m\Omega_{m-1}}{\tilde \omega}\cdot H\right)\\
+2y\sum_m \tilde c_m \frac{\Omega_{m-1}}{\tilde\omega}
\end{array}$\\

\vspace{2mm}\\

\hline
\hline

\end{tabular}

\vspace{4mm}
The next table, i.e., Table 3 includes inverse masses and potentials of superintegrable systems admitting three
integrals of motion   one of which is the shifts generator $P_3$. Three  of them include arbitrary functions while the
remaining one include arbitrary parameters whose standard number is equal to eight. Let us remind that the
notation $\varphi$ is used for the Euler angle.

Finally, the last is Table 4 which includes four maximally superintegrable systems. All of them include
arbitrary parameters whose number is varying from one to six.

The integrals of motion presented in Tables 2-4 are functionally independent. In additional there is a number
integrals of motion which are functionally independent with the presented ones but are linearly independent.
To find them it is sufficient to calculate the commutators of the presented operators with $P_3$.
The  number of linearly independent integrals of motion for the systems presented in Items 2,  3 and 6 of 
Table 3 is equal to
equal to 3 while the remaining items represent the systems for which this number is equal to 2.

The number of linearly independent integrals of motion  for the systems presented in Table 4
are equal to four (Item 1), five (Item 2), six
 (Item 3) and
 seven   (Item 4).

\newpage
\begin{center}Table 3.  Inverse   masses, potentials  and  integrals of motion for shift invariant superintegrable  systems\end{center}
\begin{tabular}{c c c c}
\hline
\vspace{2mm}No&$f$&$V$&\text{Integrals of motion}\\

\hline

\\

\vspace{2.0mm}1&$\frac{1}{c_1x_1+c_2x_2+2c_3\tilde r^2+c_4}$&$\frac{c_5x_1+c_6x_2+2c_7\tilde r^2+c_8}{c_1x_1+c_2x_2+2c_3\tilde r^2+c_4}$&$\begin{array}{c}P_1^2-P_2^2\\+\left((c_2x_2-c_1x_1+c_3(x_2^2-x_1^2))\cdot H\right)\\
+c_5x_1-c_6x_2+c_7(x_1^2-x_2^2),\\P_1P_2\\-\frac12\left(((2c_3+c_2)x_1+c_1x_2)\cdot H\right)\\+\frac12((2c_7+c_6)x_1+c_5x_2)\end{array}$\\

\hline
\\
2\vspace{2.0mm}&$\frac{1}{c_1 x_1+F(x_2)}$&$\frac{c_2x_1+\tilde F(x_2)}{c_1x_1
+F(x_2)}$&$\begin{array}{c}P_1^2-P_2^2+c_1x_1-\tilde F(x_2)\\+\left((F(x_2)-c_1x_1)\cdot H\right),\\
 \{P_3,L_2\}-\frac{c_1}2(x_3^2\cdot H)+\frac12c_2x_3^2
\end{array}$\vspace{2mm}\\

\hline
\\
3\vspace{2.0mm}&$\frac{x_2^2}{x_2^2F(x_1)+c_1}$&$\frac{x_2^2\tilde F(x_1)+c_2}{x_2^2F(x_1)+c_1}$
&$\begin{array}{l}L_1^2-\left(\frac{c_1x_3^2}{x_2^2}\cdot H\right)+\frac{c_2x_3^2}{x_2^2},\\P_1^2-P_2^2+\left((\frac{c_1}{x_2^2}-F(x_1))\cdot H\right)\\-\frac{c_2}{x_2^2}+\tilde F(x_1)
\end{array}$\\

\hline

\\

4\vspace{2mm}&$\frac{x_1^2x_2^2}{c_1x_1^2+c_2x_2^2+c_3x_1^2x_2^2\tilde r^2+c_4x_1^2x_2^2}$&$\frac{c_5x_1^2+c_6x_2^2+c_7x_1^2x_2^2\tilde r^2+c_8x_1^2x_2^2}{c_1x_1^2+c_2x_2^2+c_3x_1^2x_2^2\tilde r^2+c_4x_1^2x_2^2}$&$\begin{array}{c}L_3^2-\left(\frac{c_1x_1^4+c_2x_2^4}{x_1^2x_2^2}\cdot H\right)+\frac{c_5x_1^4+c_6x_2^4}{x_1^2x_2^2},\\P_1^2-P_2^2\\+\left(\frac{c_1^2x_2^2-c_2^2x_2^2
-c_3(x_1^2-x_2^2)x_1^2x_2^2}{x_1^2x_2^2}\cdot H\right)\\-\frac{c_5^2x_2^2-c_6^2x_2^2
-c_7(x_1^2-x_2^2)x_1^2x_2^2}{x_1^2x_2^2}\end{array}$\\

\hline
\\
5\vspace{2mm}&$\frac{x_1^2}{c_1(x_1^2+4x_2^2)x_1^2+c_2x_1^2x_2+c_3+c_4x_1^2}$
&$\frac{c_5(x_1^2+4x_2^2)x_1^2+c_6x_1^2x_2+c_7+c_8x_1^2}{c_1(x_1^2+4x_2^2)x_1^2+c_2x_1^2x_2+c_3+c_4x_1^2}$&
$\begin{array}{c}P_1^2-P_2^2-c_6x_2+\frac{c_7}{x_1^2}\\+\left((c_1(4x_2^2-x_1^2)+c_2x_2)\cdot H\right)\\-\left(\frac{c_3}{x_1^2}\cdot H\right)-c_5(4x_2^2-x_1^2),\\
\{P_1,L_3\}+\frac{4x_2(c_5x_1^4-c_7)+c_5x_1^4}{2x_1^2}\\-\left(\frac{4x_2(c_1x_1^4-c_3)+c_1x_1^4}{2x_1^2}\cdot H\right)
\end{array}$\\

\hline
\\

6\vspace{2mm}&$\frac{\tilde r^2 }{c_1\tilde r^2+F(\varphi)}$&$\frac{c_2\tilde r^2
+G(\varphi) }{c_1\tilde r^2+F(\varphi)}$&$\begin{array}{c}L_3^2-(F(\varphi)\cdot H)-G(\varphi),\\D^2-c_1(r^2\cdot H)+c_2r^2\end{array}$\\

\hline
\\
7\vspace{2mm}&$\frac{\tilde r^2}{c_1(\mu\ln(\tilde r)-\nu\varphi)+c_2\tilde r^2+c_3}$&
$\frac{c_4(\mu\ln(\tilde r)-\nu\varphi)+c_5\tilde r^2+c_6}{c_1(\mu\ln(\tilde r)-\nu\varphi)+c_2\tilde r^2+c_3}$&
$\begin{array}{c}\mu P_3L_3+\nu\{P_3,D\}\\
-\left(2c_2\nu x_3\cdot H\right)+2c_5\nu x_3,\\ L_3^2+2(c_1\nu\varphi\cdot H)-2c_4\nu\varphi\end{array}$\\

\hline
\\
8\vspace{2mm}&$\frac{\tilde r x_1^2}{c_1x_1^2+c_2x_2+c_3\tilde r x_1^2+c_4\tilde r}$&
$\frac{c_5x_1^2+c_6x_2+c_7\tilde r x_1^2+c_8\tilde r}{c_1x_1^2+c_2x_2+c_3\tilde r x_1^2+c_4\tilde r}$&
$\begin{array}{c}L_3^2-\left(\frac{c_2\tilde r x_2+c_4x_2^2}{x_1^2}\cdot H\right)
-\frac{c_6\tilde r x_2+c_8x_2^2}{x_1^2},\\\{P_1,L_3\}-\frac{2c_8x_2\tilde r+c_6(x_1^2+2x_2^2)+c_5x_2x_1^2}{x_1^2\tilde r}\\+\left(\frac{2c_4x_2\tilde r+c_2(x_1^2+2x_2^2)+c_1x_2x_1^2}{x_1^2\tilde r}\cdot H\right)\end{array}$\\

\hline
\hline

\end{tabular}

\vspace{2mm}

\begin{center}Table 4.  Inverse   masses, potentials  and  integrals of motion for shift invariant maximally superintegrable  systems\end{center}
\begin{tabular}{c c c c}
\hline

\vspace{1.5mm}No&$f$&$V$&\text{Integrals of motion}\\
\hline
\\
1\vspace{1.5mm}&$\frac{1}{ x_1}$&$\frac{c_1x_2}{ x_1}$&
$\begin{array}{l}P_1^2-P_2^2-c_1x_2-\left(x_1\cdot H\right),\\
2P_1P_2+c_1x_1-(x_2)\cdot H),
\\ \{P_3,L_2\}-\frac{1}2(x_3^2\cdot H),\
\{P_3,L_1\}-\frac{c_1x_3^2}2
\end{array}$\\

\hline
\\

2\vspace{1.5mm}&$\frac{x_1^2x_2^2}{c_1x_1^2+c_2x_2^2+c_3x_1^2x_2^2}$&
$\frac{c_4x_1^2+c_5x_2^2+c_6x_1^2x_2^2}{c_1x_1^2+c_2x_2^2+c_3x_1^2x_2^2}$&
$\begin{array}{l}L_3^2-\left(\frac{c_1x_1^4+c_2x_2^4}{x_1^2x_2^2}\cdot H\right)+
\frac{c_5x_1^4+c_6x_2^4}{x_1^2x_2^2},\\
L_1^2-\left(\frac{c_1x_3^2}{x_2^2}\cdot H\right)+\frac{c_4x_3^2}{x_2^2},
\\D^2-\left(c_3x_3r^2\cdot H\right)+c_6x_3r^2\end{array}$
\\
\hline
\\

3\vspace{1.5mm}&$\frac{\tilde r x_1^2}{c_1x_2+c_2\tilde r+c_3\tilde r x_1^2}$&
$\frac{c_4x_2+c_5\tilde r+c_6\tilde r x_1^2}{c_1x_2+c_2\tilde r+ c_3\tilde r x_1^2}$&$\begin{array}{l}L_3^2
-\left(\frac{c_1\tilde r x_2+c_2x_2^2}{x_1^2}\cdot H\right)
-\frac{c_4\tilde r x_2+c_5x_2^2}{x_1^2},
\\
D^2-\left(c_2r^2\cdot H\right)+c_5r^2,\\
\{L_2,L_3\}-\frac{2c_5x_3\tilde r+c_4x_3(x_1^2+2x_2^2)}{x_1^2\tilde r}\\
+\left(\frac{2c_2x_3\tilde r+c_1x_3(x_1^2+2x_2^2)}{x_1^2\tilde r}\cdot H\right)
\end{array}$\\

\hline
\\

4\vspace{1.5mm}&$\frac{ x_1^2}{x_2x_1^2+c_1}$&$\frac{c_2 x_2x_1^2+c_3+c_4x_1^2}{x_2x_1^2+c_1}$&
$\begin{array}{l}L_2^2-\left(\frac{c_1x_3^2}{x_1^2}\cdot H\right)+\frac{c_3x_3^2}{x_1^2},
\\\{P_2,D\}-\frac12\left((x_1^2+4x_2^2+x_3^2)\cdot H\right)\\+\frac{c_2}2(x_1^2+4x_2^2+x_3^2)+2c_4x_2,\\
\{P_3,L_1\}+\left(\frac{c_1x_3^2}2\cdot H\right)-\frac12c_3x_3^2
\end{array}$\\

\hline\hline
\end{tabular}

\vspace{2mm}

\section{Discussion}
We classify inequivalent quantum mechanical systems with position dependent masses which admit
second order integrals of motion and  one out of two  one parametric Lie  groups, i.e., dilatation and shift ones. The total number of such
systems is equal to twenty seven, including nine ones invariant w.r.t. the dilatation
transformations and eighteen shift invariant  systems. The classification results are presented in four
 tables and include eleven integrable, ten superintegrable and six maximally superintegrable systems.
 Let us note that this statement is a bit conventional since the obtained systems are defined up to arbitrary
 functions and (or) arbitrary parameters, and in fact we enumerate the  subclasses of the systems under
 study including the infinite sets of them. One more note is that the systems represented in Item 5 of Table 1
 and Item 8 of Table 2 form the subclasses of the systems represented in Item 4 of Table 1 and Item 7
 of Table 2 correspondingly.

The presented  results  can be treated as the next step in our programm of the complete classification of the
3d PDM systems admitting second order integrals of motion. Such systems possessing the symmetry w.r.t.
three and two parametric Lie groups were classified in paper \cite{AG} and \cite{AG1} respectively.
The next step presupposes the classification of the systems admitting at least one parametric Lie symmetry
group. In accordance with \cite{NZ}, up to equivalence  it is sufficient to restrict ourselves to the two
subclassess of the groups. The first of them  includes  rotations around the fixed axis, shifts along this axis
and dilatations, we call these symmetries natural. The second subclass includes the so called combined
transformations and includes superpositions of shifts and rotations, dilatations and rotations,  and  shift,
rotations and conformal transformations.

The main result of the present paper consists in completing the classification of the integrable
systems admitting the natural symmetries started in paper \cite{AGG}. Thus to complete the classification of the 3d PDM systems admitting second order integrals of motion and a one parametric Lie symmetry
group it is sufficient to classify the the systems admitting the combined symmetries mentioned in the above. This work is in progress.

One out of many stimulus to search for integrable and superintegrable hamiltonian systems consists in the fact
that such systems as a rule are exactly solvable. This statement is an element of common knowledge for
standard quantum mechanical systems. However, the same is true for systems  with position dependent masses.
In particular, it is the case for the rotationally invariant superintegrable PDM systems. It was shown in
paper   \cite{154} that all such systems are exactly solvable. Moreover, in addition to the superintegrability,
they are shape invariant with respect to the Darboux transform and can be effectively solved using the tools of
supersymmetric quantum mechanics.   Exact solvability of the PDM systems with extended Lie symmetries  was prowed
in paper \cite{AGno}.

The testing for exact solvability of superintegrable PDM systems lies out of frame of the present paper. However,
it is possible to declare that at least some of the found systems do are separable and exactly solvable. In particular,
it is the case for the system presented in Item 1 of Table 4. The  study of the exact solvability aspects of the
classified systems is one more problem which we plane to consider in  the future research.

{\bf Acknowledgement.} I am indebted with Universit\'a del Piemonte Orientale and and Dipartimento di Scienze e Innovazione Tecnologica for the extended stay as Visiting Professor.

\section{Appendix}

In our previous papers, e.g., in  \cite{AG5} we   following form of hamiltonian (\ref{A1}):
\begin{gather}\la{Ha} H=p_afp_a+ \hat V.\end{gather}

Operators (\ref{Ha}) and (\ref{H}) are equal one to another provided potentials $V=V({\bf x})$ and $\hat V=\hat V({\bf x})$ satisfy the following condition:
\begin{gather}\la{Hb}\hat V =V-V^k\end{gather}
where
\begin{gather}\la{Hc}V^{(k)}=\frac1{4f}((\p_1 f)^2+(\p_2 f)^2+
(\p_3 f)^2)
-\frac12\Delta f\end{gather}
and $\Delta$ is the Laplace operator.

Formula (\ref{Hc}) represents an example of {\it kinematical} potentials which can be reduced to zero by
the rearranging  the   ambiguity parameters. One of the typical properties of such potentials is their independence
on coupling constants.

The form (\ref{Ha}) of the Hamiltonian is more nice than (\ref{H}), however the corresponding potential is not
necessary free of the kinematical part. On the other hand  Hamiltonian in form (\ref{H}) does not include such
part. In addition, the latter form is more convenient for using the St\"ackel transform with can be made by the
simultaneous
multiplication of this operator by the square roots of the potential  from the l.h.s. and r.h.s. And it is why we use
just realization  (\ref{H}) in the present paper.

Just representation (\ref{Hb}) for the PDM Hamiltonian was used in \cite{154}
where rotationally invariant systems were described. The results presented in \cite{AG, AG1, AG5}
are valid for the representation  (\ref{Hb}) also,  but not for representation  (\ref{Ha}). We present the
requested comments in the preprint versions of the mentioned papers.

\end{document}